\documentclass[preprint]{aastex631}
\usepackage{hyperref}
\usepackage{graphicx}
\usepackage[caption = false]{subfig}
\usepackage{amsmath, amssymb}
\usepackage{float} 

\newcommand{\target}{FRB~20240114A}

\begin{document}

\title{A flaring radio counterpart to a fast radio burst reveals a newborn magnetized engine}

\correspondingauthor{Wenfei Yu, Bing Zhang}
\email{wenfei@shao.ac.cn, bing.zhang@unlv.edu}
%\correspondingauthor{Bing Zhang}
%\email{bing.zhang@unlv.edu}

\author[0000-0002-8086-4049]{Xian Zhang}
\affiliation{Shanghai Astronomical Observatory, Chinese Academy of Sciences, Shanghai 200030, China}

\author[0000-0002-3844-9677]{Wenfei Yu}
\affiliation{Shanghai Astronomical Observatory, Chinese Academy of Sciences, Shanghai 200030, China}

\author[0000-0002-5385-9586]{Zhen Yan}
\affiliation{Shanghai Astronomical Observatory, Chinese Academy of Sciences, Shanghai 200030, China}

\author{Yi Xing}
\affiliation{Shanghai Astronomical Observatory, Chinese Academy of Sciences, Shanghai 200030, China}

\author[0000-0002-9725-2524]{Bing Zhang}
\affiliation{The Nevada Center for Astrophysics, University of Nevada, Las Vegas, Las Vegas, NV 89154, USA}
\affiliation{Department of Physics and Astronomy, University of Nevada, Las Vegas, Las Vegas, NV 89154, USA}



\begin{abstract}

Fast Radio Bursts (FRBs) are energetic millisecond radio bursts at cosmological distances, whose underlying engine is not identified. Among a sub-population that emit repeated radio bursts, a handful were associated with a persistent radio source (PRS) whose origin is unknown. Here we report the discovery of a compact flaring radio source (FRS) associated with a newly-active repeating FRB within one month after the first radio burst was detected. Its temporal and spectral characteristics differ from those of the PRSs but are similar to those of engine-powered supernovae and low-luminosity active galactic nuclei. We detected a spectral peak around $1.6\pm0.2$ GHz that is consistent with synchrotron self-absorption. Assuming equipartition, the magnetic field strength in the FRS is larger than the line-of-sight component constrained from the FRB Faraday rotation, suggesting a highly magnetized engine. The radius of the FRS is constrained to be $\sim0.03$ pc and the minimum total energy is $\sim~6.2\times{10}^{47}~{\rm ergs~}$. This FRS reveals the birth of a highly magnetized FRB engine, and hints that PRSs associated with other active FRBs may be the later stage of evolution.

\end{abstract}



\section{Introduction} \label{sec:intro}
Fast Radio Bursts (FRBs) are bright millisecond-duration radio bursts at cosmological distances \citep{2007Sci...318..777L,2013Sci...341...53T} with unknown origin. At least some FRBs are powered by magnetars \citep{CHIME-SGR,STARE2-SGR,HXMT-SGR,Integral-SGR}, but other engines, including pulsars, stellar-sized or even supermassive black holes, remain possible to interpret some FRBs (\citealt{zhang23} and references therein).  
%BZ: Sorry I deleted these references. There are many magnetar references and pulsar references, and it is unclear why the two Sridhar papers attract more attention.
%Although still mysterious, they are thought powered by compact objects, such as those neutron stars and/or black holes in magnetars \cite{2013arXiv1307.4924P}, pulsars \cite{2016ApJ...829...27D}, X-ray Binaries (XRBs)~\cite{2021ApJ...917...13S,2022ApJ_Sridhar}, and Active Galactic Nucleus (AGN). 
Persistent or variable multi-wavelength/multi-messenger counterparts to FRBs are essential to unveil the underlying engine of FRBs, but most searches have led to null results \citep{LVK23,curtin23,pearlman24,zhang24}.

%, as potential counterparts would allow direct probes of their local environment, formation history, and origin of the FRB engine. However, most of the efforts have not been successful\cite{2015MNRAS.447..246P,2022arXiv220312038T}. 

So far, the only credible electromagnetic counterpart for cosmological FRBs is the so-called ``persistent radio source'' (PRS) that was found to be associated with a handful of actively repeating FRB sources \citep{2017Natur.541...58C,2022Nature_Niu,2024Natur.632.1014B,2024ApJ...976..199I}. Among them, FRB 20121102A and FRB 20190520B were identified to have compact PRS sizes with a radius $\rm R<5$ pc \citep{2017ApJ...834L...8M,2023ApJ...958L..19B}. Both sources show substantial local contribution to the dispersion measure (DM) \citep{2017Natur.541...58C, 2022Nature_Niu}, extremely large Faraday Rotation Measure (RM)  \citep{2018Natur.553..182M,2022Sci...375.1266F}, and long-term RM variations and even sign reversals \citep{2023Sci...380..599A}. %quantitatively distinct from other repeating and non-repeating FRBs. 
%BZ: We may want to speculate the following, because our new-born, most active source has a lower RM. 
%They might represent either a different evolution phase or a different formation scenario due to their dense and dynamic magneto-ionic near-source environment \cite{2018Natur.553..182M,2022Nature_Niu}. 
A third, less compact ($\sim$700 pc) PRS and two other candidates with lower RM values have been identified later \citep{2024Natur.632.1014B,2024ApJ...976..199I}. In general, it is believed that these PRSs are synchrotron-emitting nebula around FRB sources \citep{2020ApJ...895....7Y,2022ApJ...928L..16Y}, possibly a magnetar wind nebula in a supernova remnant \citep{2016ApJ...819L..12Y,2016MNRAS.461.1498M,2017ApJ...841...14M,2018ApJ...868L...4M,2021ApJ...923L..17Z} or a hypernebula associated with an accreting stellar-mass black hole \citep{2022ApJ_Sridhar}. The synchrotron nature of PRSs is supported by the predicted correlation between PRS luminosity and RM that has been verified by the growing observations \citep{2020ApJ...895....7Y,2022ApJ...928L..16Y,2024Natur.632.1014B,2024ApJ...976..199I}. 

The formation and evolution of a PRS have not been observationally studied so far. The two compact PRSs were discovered about 1.2 \citep{2022Nature_Niu,2023ApJ...959...89Z} and 3.8 years \citep{2017Natur.541...58C} after the discovery of the corresponding FRB; the third PRS was first detected 10 months after its burst discovery \citep{2024Natur.632.1014B}. These PRSs are at least 3--12 years old now and do not show strong evolution. But a synchrotron nebula of pc sizes or above would need years to form. Observations in the earlier phases of the PRS emergence are desired to understand the origin of PRSs and the FRB sources.

\section{The new, hyper-active \target}

\target~is a repeating FRB discovered by the CHIME/FRB Collaboration on 2024 January 14 \citep{2024ATel16420....1S}. Radio observations with multiple telescopes across the globe have detected hundreds of bursts or more from 300 MHz up to 6~GHz, showing hyperactivity in the form of episodes of high burst rates \citep{2024ATel16432....1O,2024ATel16452....1K,2024ATel16505....1Z,2024ATel16597....1H,2024ATel16620....1L}. Precise localizations achieved with the MeerKAT \citep{2024ATel16446....1T} and the European VLBI Network (EVN) \citep{2024ATel16542....1S} observations helped the identification of its host galaxy J212739.84$+$041945.8 at the redshift of $z$ = 0.13 \citep{2024ATel16613....1B}. Efforts to search for X-ray \citep{2024ATel16645....1V} and radio  counterparts \citep{2024ATel16452....1K,2024arXiv240509749P,2024arXiv240612804K} only gave upper limits. Coincident gamma-ray activities in the form of gamma-ray flares with a duration of a few tens of seconds up to the gigavolt energy and a peak luminosity of ${10}^{48}~{\rm ergs}~{\rm s}^{-1}$ have been discovered in the direction of the FRB within 0.3 degrees \citep{2024ATel16594....1X,2024ATel16630....1X,2024arXiv241106996X}. The unprecedented hyper-active radio-bursting behaviour and coincident energetic gamma-ray flaring activities suggest a powerful central engine.

\section{A flaring radio source associated with \target}

%The localization of \target~at the arcsecond level was achieved by the MeerKAT observations performed on 2024 February 9 \cite{2024MNRAS.533.3174T}. The observation consists of an L-band observation of a total exposure time of $\sim$ 1.4 hours and an on-source time of $\sim$ 1 hour. The visibilities of 2 seconds sampling time were saved for the observations (see Appendix). By engaging complex flagging and calibration processes and careful analysis, we were able to obtain a clear image of the FRB field. 

We discovered a compact Flaring Radio Source (FRS) in spatial association with the FRB with MeerKAT and the Karl G. Jansky Very Large Array (VLA) observations, which were performed as early as only 26 days after its burst discovery \citep{2024ATel16695....1Z} (see below). The source was also detected later with the upgraded Giant Metrewave Radio Telescope (uGMRT) at lower frequencies \citep{2024ATel16820....1B,2024arXiv241213121B} and with the Very Long Baseline Array at 5 GHz 
%with observations up to 2024 September 
\citep{2024arXiv241201478B}, further supporting the association of the FRS with \target. 

%We discovered the radio continuum counterpart with those MeerKAT observations which were performed only 26 days after the FRB discovery 
We detected the source at the position of (R.A., Decl.)~[J2000] = (21h27m39.82s, 4\arcdeg19\arcmin47.106\arcsec) with 1 $\sigma$ statistical uncertainty of (0.65\arcsec, 0.89\arcsec) \citep{2024ATel16695....1Z} with the MeerKAT observations on 2024 February 9 (Figure \ref{fig:skyim}, and see Appendix). Its flux density was 72$\pm$14 $\mu$Jy at 1.3~GHz, corresponding to a spectral luminosity of $L_{\rm{1.3~GHz}}=3.2\times10^{28}$\ erg s$^{-1}$\ Hz$^{-1}$; the in-band (0.86--1.71~GHz) spectral index was -1.1$\pm$0.8. A 3 $\sigma$ upper limit of 120 $\mu$Jy/beam at the central frequency of 812~MHz was achieved with the accompanied UHF-band observation. 

We confirmed the detection of the source and its steep radio spectra with VLA Director's Discretionary Time (DDT) multi-band observations performed in 2024 July 23--29 and afterward (Figure \ref{fig:skyim}, and see Appendix). The position of the source is constrained more accurately at (R.A., Decl.)~[J2000] = (21h27m39.84s, 4\arcdeg19\arcmin45.598\arcsec) with 1 $\sigma$ statistical uncertainty of (0.06\arcsec, 0.04\arcsec) from the C-band deep image. The flux density in L-band (1.5 GHz) had increased from 64$\pm$14 $\mu$Jy at 1.5 GHz measured on February 9 to 131$\pm$10 $\mu$Jy measured in late July (Figure \ref{fig:lc_spec}). The peak spectral luminosity corresponds to $L_{\rm{1.5~GHz}}=5.8\times10^{28}$\ erg s$^{-1}$\ Hz$^{-1}$. Such a $\sim$100\% increase in flux density indicates a rising radio flare. Together with the accompanied broadband spectral variation and steeper spectra, we define this radio counterpart as an FRS (see Appendix), in contrast PRSs that show stable spectra and less variability \citep{2023ApJ...959...89Z,2024ApJ...976..165Y}. In addition to the rise, we also found an apparent, broadband decline in flux densities of the FRS since late July in the VLA observations performed on October 3, with the L-band (1.5 GHz) flux detected at 87$\pm$19 $\mu$Jy. 

The corresponding e-folding rise and decline time scales are $\tau_{r} = 235\pm76$ days and $\tau_{d} = 168\pm95$ days, respectively, implying a compact nature of the source; the corresponding light-crossing radius for the rise is only $\sim$0.18$\delta$ pc (where $\delta$ is the Doppler, see Appendix). The size constraint of the FRS is smaller than the best constraint (${\rm R} \sim 0.35~{\rm pc}$, $\delta=1$ assumed) of the PRSs \citep{2017Natur.541...58C}.

\section{Comparison with various types of slow radio transients}

The radio flux density measurements of the FRS are shown in Figure \ref{fig:lc_spec}. There were no radio interferometry observations sensitive enough to cover the likely rise until our VLA observations performed $\sim$170 days after the MeerKAT observations. The actual peak of the flux density could have reached before the VLA observations. Therefore, the observed peak flux density measured in 2024 July should be regarded as a lower limit and the rise time scale is likely shorter. Radio flares have been observed in a wide range of astronomical objects such as supernovae (SNe), gamma-ray bursts (GRBs), tidal disruption events (TDEs), X-ray binaries (XRBs), active galactic nuclei (AGN), and blazars. Their peak luminosity and the characteristic e-folding rise or decay timescales can be used to diagnose source origins \citep{2015MNRAS.446.3687P}. The FRS falls in the most luminous end of the SNe sample and the dimmest end of the low-luminosity AGN (LLAGN) sample (Figure \ref{fig:comparison_transients_w} and Figure \ref{fig:comparison_transients_efolding}, see also Appendix). Among the radio SNe observed to date, only SN Ib/c and SN IIn types can reach the same peak spectral luminosity; but the FRS seems to have characteristic rise or decay timescales longer than those of SN Ib/c but shorter than those of SN IIn \citep{2019ApJ...876L..10E}. We also found that the FRS tends to join the engine-powered SNe among the radio SNe \citep{2012ApJ...752...78S} (see Figure \ref{fig:FRS_sne}). On the other hand, the FRS has a lower peak luminosity and a shorter variability timescale as compared to those radio flares in LLAGNs. A radio flare of an accreting black hole with a lower mass than those of AGN might behave this way. The similarity of the FRS to engine-powered SNe and low-luminosity AGN suggests that the FRB engine may invoke both explosive and accretion/jet processes. 

\section{The evolving radio spectra and synchrotron self-absorption}

The broadband radio spectrum of the FRS evolved from a power-law with a spectral index -1.1$\pm$0.8 in the L-band, with a likely low-frequency cut-off due to the upper limit of the UHF-band in 2024 February, to a broken power law with a prominent spectral peak in 1--2 GHz in 2024 July, and then to a flatter power law with a spectral index of -0.43$\pm$0.21 measured in early October (Figure \ref{fig:spectral_evolution}). Specifically, the broadband radio spectrum measured by the uGRMT and the VLA observations in the period between June and August showed a prominent spectral peak at $\nu_{b}$=1.56$\pm$0.20 GHz (Figure \ref{fig:lc_spec}), which is consistent with a synchrotron self-absorption (SSA) spectral break rather than other mechanisms (see Appendix). The peak flux density obtained at $\nu_{b}$ from the best-fit model is 145.8$\pm$14.5 $\mu$Jy. Below and above $\nu_{b}$, the spectral indices are 0.91$\pm$0.27 and -0.73$\pm$0.14, respectively. The SSA spectral break frequency, corresponding to the opacity $\tau=1$, is at $\nu_{\tau=1}= 1.41 \pm 0.18$ GHz (see Appendix). Since the radio bursts from the engine need to get through the FRS to reach the observer, those bursts with an emission frequency below $\nu_{\tau=1}$ could be absorbed and used to heat the plasma \citep{2016ApJ...819L..12Y}, as $\nu_{\tau=1}$ sets the boundary of the optically-thick and optically-thin regimes. The frequency-dependent burst activities observed by CHIME, FAST, and other telescopes might be the result of the modulation due to an evolving $\nu_{\tau=1}$, which causes the high burst rate episodes seen with CHIME, FAST, and other telescopes at higher frequencies as being not simultaneous. 

\section{The extreme properties of the flaring radio source}

The detection of the SSA spectral break indicates the radio-emitting plasma of the FRS is dense and magnetized. According to the SSA theory, the relation between the radius $R$ and the magnetic field strength $B$ of the plasma can be constrained with the $\nu_{\tau=1}$ and the flux density at $\nu_{\tau=1}$ measured from its broadband radio spectrum \citep{essential2016} (see Appendix). The corresponding $B$--$R$ relation for the FRS is shown as shaded orange in Figure \ref{fig:B_R_constraints}. Assuming equipartition between the electron energy and the magnetic energy, one can derive another constraint as indicated by green solid lines. The joint constraints led to the constraint of the magnetic field strength of $B_{\rm eq} \sim 0.054 \pm 0.007$ G and the size $R_{\rm eq} \sim 0.027 \pm 0.004$ pc  of FRS, respectively. The nondetection of an SSA frequency down to hundreds of MHz in the PRS of FRB~20121102A constrains its parameter regime in the pink region \citep{2021A&A...655A.102R}. One can see that the FRS of \target~has a much smaller size while carrying a stronger magnetic field, suggesting that the FRS may be either of a different physical origin or at an earlier stage of a PRS. Assuming equipartition, one can also estimate the minimum energy of the plasma based on the magnetic energy to be $\sim 6.2\times10^{47}~{\rm ergs}$. With a radius of 0.027 pc, the brightness temperature of the plasma during the 2024 July observations was $4.7\times{10}^{11}$ K, more than two or three orders of magnitude higher than those ever achieved for any PRSs. This is close to maximum brightness temperature $T_{\rm max} \sim \gamma_e m_e c^2 / k \sim (5.9\times 10^{11}) \gamma_{e,2}$ K of a non-relativistic nebula with typical electron Lorentz factor $\gamma_e \sim 100 \gamma_{e,2}$ \citep{zhang23}.  

Estimate of the magnetic field strength along the Line-of-Sight (LoS) may be derived using the relation $B_\parallel= (1.23\times{10}^{-6} \ {\rm G}) \frac{\rm RM_{src}}{\rm DM_{src}}$ from the $\rm RM_{src}$ and $\rm DM_{src}$ of the FRB, where the sub-script ``src'' stands for the contribution from the ``source'', i.e. the immediate environment of the FRB, which is mainly contributed by the FRS. The measured RM is $338.1\pm0.1~{\rm rad}~{\rm m}^{-2}$, which corresponds to a local (redshift-corrected) RM of $\sim (449\pm23)~{\rm rad}~{\rm m}^{-2}$ \citep{2024MNRAS.533.3174T}. This value can be regarded as $\rm RM_{src}$. On the other hand, $\rm DM_{src}$ can be constrained using two independent methods. First, the measured DM is $527.65\pm0.01~{\rm pc}~{\rm cm}^{-3}$. The DM contribution from the host galaxy can be estimated as ${333}^{+90}_{-125}~{\rm pc}~{\rm cm}^{-3}$ \citep{2024MNRAS.533.3174T}, and the corresponding rest-frame value is $\sim {376.3}~{\rm pc}~{\rm cm}^{-3}$. Taking this as the maximum $\rm DM_{src}$, one can derive $B_\parallel > 1.5\times 10^{-6}$ G. This is more than four orders of magnitude smaller than $B_{\rm eq}$. Second, a conservative lower limit of $\rm DM_{src}$ can be derived from the FRS observations. Using the minimum plasma energy derived based on the magnetic equipartition condition, the electron density derived from the brightness temperature, and the volume derived from the source size, one can derive ${\rm DM_{src}} \geq 0.035 ~{\rm pc}~{\rm cm}^{-3}$. This gives an upper limit of the magnetic field strength along the LoS, i.e. $B_{\parallel} \leq 1.6\times{10}^{-2}$ G, which is more than a factor of 3.4 smaller than $B_{\rm eq}$.

%The conclusion that the total magnetic field $B$ is significantly larger than $B_\parallel$ is robust for a deviation from equipartition toward an increasing electron energy density $\epsilon_{e}$, as it is accompanied with a decreasing $B$ and will increase the electron density and the corresponding DM by a larger factor (see Appendix). 

If $B_\parallel$ is close to the upper bound, then one can reach a consistent estimate of $B$ in the nebula assuming that the magnetic energy is somewhat below equipartition. However, if $B_\parallel$ is close to the lower bound, then one runs into the large discrepancy between the two $B$ estimates. 
%The discrepancy between the two values of magnetic fields ($B_{\rm eq}$ vs. upper limit on $B_\parallel$) shed light onto the composition of the FRS plasma. 
The simplest interpretation is under the assumptions that the plasma is an ion plasma and that both $B$ estimates are from the same radio-emitting region. In this case, the magnetic field configuration must be dominated by a component that is perpendicular to the LoS. This is consistent with a Poynting-flux-dominated outflow moving toward Earth, in which toroidal component of magnetic field (which falls with radius as $B_{\rm t} \propto r^{-1}$) dominates over the poloidal component (which falls with radius as $B_{\rm p} \propto r^{-2}$). In the second scenario, the FRS emitter is electron-positron pair dominated, likely injected from a highly magnetized central engine. The large magnetic field strength in the emission region does not introduce net RM \citep{yang23}. The $\rm RM_{src}$ is provided by an ion plasma in the foreground of the FRB source where magnetic field is much weaker. 
%BZ: I tend to discard this one. Does not help us to make the case. 
%In the third scenario, the FRS is patchy and does not cover the LOS. However, the large contrast in $B$ measurements make this scenario unlikely. 
In either case, the plasma contributing to the FRS emission is likely associated with fresh injection from a newborn magnetized engine.

\section{Evidence of a newborn engine and implications}

The observations of FRB~20240114A and its associated emissions point toward a consistent picture of a newborn FRB engine. Evidence in support of this conclusion includes: (1) The source was newly discovered and is extremely active, with the highest burst rate detected by multiple telescopes; (2) The source was associated with short $\gamma$-ray flares detected by the Fermi LAT \citep{2024arXiv241106996X}; (3) The associated continuous radio emission shows a flaring behavior with an evolving spectrum, in contrast to the stable spectrum and luminosity of PRSs associated with other FRBs, but similar to SNe, especially engine-powered ones; (4) The size of the FRS is much smaller than other PRSs and the magnetic field much stronger, consistent with a Poynting-flux-dominated outflow from a newborn engine; (5) The discrepancy between $B_{\rm eq}$ and $B_\parallel$ may suggest a Poynting-flux-dominated outflow. 

\begin{figure}[H]
 \centering
 \includegraphics[angle=0,width=6.0in]{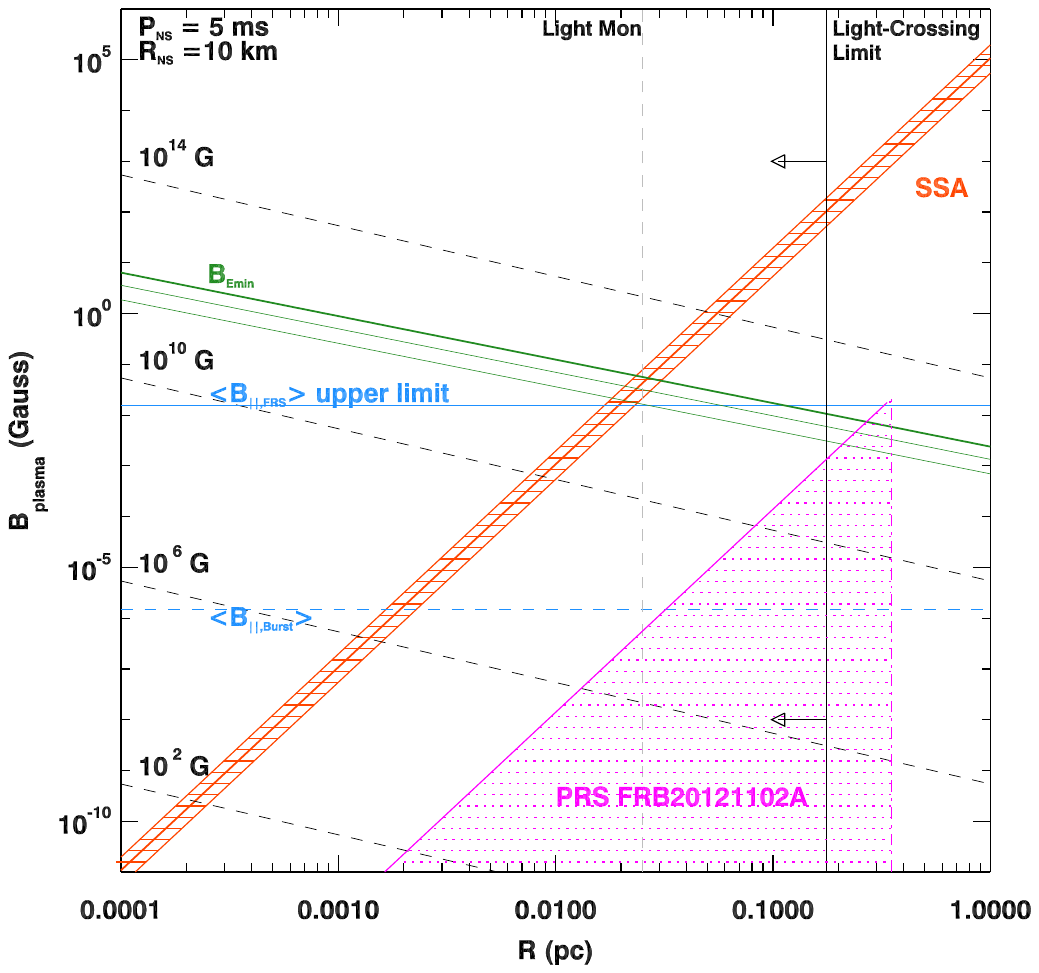}
 %{mysynchrotron_selfabsorption_4frb.pdf}
 \vspace{-3.cm}
 \caption{
 \footnotesize \textbf{Constraints on the magnetic field strength ${\rm B}$ and the source radius ${\rm R}$ for the FRS.} The allowed regime by the SSA constraints is shaded in orange. The regime defined by an equipartition (green line) with the deviations by a factor of 10 and 100 (thin green lines) are also shown. Their intersection corresponds to magnetic field strength $B_{\rm eq}=0.054\pm0.007$ G and radius $R_{\rm eq}=0.027\pm0.004$ pc. The magnetic field along the line-of-sight estimated for the FRS ${\rm B}_{\parallel,\rm FRS}$ ($\leq$ 0.016 G, solid) and for the bursts ${\rm B}_{\parallel,{\rm Burst}}$ ($> 1.5\times{10}^{-6}$ G, dashed) are shown as blue horizontal lines (see Appendix). The black dashed lines show the corresponding $B\propto{R}^{-1}$ relation outside the light-cylinder from the neutron star with the corresponding dipole magnetic field, a spin period of 5 ms, and a radius of 10 km. The solid vertical line with arrows pointing toward smaller radii marks the upper limit on the source radius (0.18 pc, $\delta=1$) from light-crossing argument for the flare rise. The radius corresponding to a light month is plotted as a vertical dashed line in gray, close to $R_{\rm eq}$. The regime for the non-detection of SSA in the PRS of FRB~20121102A \citep{2021A&A...655A.102R} is shaded in magenta.}
 \label{fig:B_R_constraints}
\end{figure}

The widely discussed engine of FRBs is magnetars, which has been supported by the observations of the Galactic FRB 20200428D \citep{CHIME-SGR,STARE2-SGR,HXMT-SGR,Integral-SGR}. It has been suggested that a new-born magnetar is likely associated with the birth of a magnetar-powered nebula \citep{2016ApJ...819L..12Y,2016MNRAS.461.1498M,2017ApJ...841...14M,2017ApJ...839L...3K,2024arXiv241219358B}. Our data is generally consistent with such a physical picture. Assuming an initial period of $P_{\rm NS} = 5$ ms for a newborn neutron star and considering $B \propto r^{-3}$ and $B \propto r^{-1}$ inside and outside the light cylinder, respectively, one can estimate the magnetic field strength of the pulsar wind at the FRS radius. To match the observations, we estimate the surface magnetic field strength of the newborn neutron star to be on the order of ${10}^{12}$ G (see Figure \ref{fig:B_R_constraints}; also Appendix). This is below the magnetar range, but consistent with the range required to account for the $\gamma$-ray emission \citep{2024arXiv241106996X}. It is known that newborn neutron stars can have strong toroidal magnetic fields so that they can still behave as magnetars. So this source can serve as the engine of the very active FRB 20240114A. Other possible engines of cosmological FRBs include stellar-mass or even intermediate-mass black holes, whose accretion would power relativistic jets similar to AGN/blazars of supermassive black holes. For a black hole with a mass $M_{\rm BH}$, to match the observations, we estimate the helix magnetic field at the jet base at $6 \ {r}_{g}$ ($r_{ g}=GM_{\rm BH}/c^2$) is on the order of $5\times{10}^{9}~{\rm G}~~{(\frac{M_{\rm BH}}{M_{\odot}})}^{-1}$ (see Appendix). 

Regardless of its identity, this engine is continuously injecting energy to the remnant, as evidenced by the non-stop emission of FRB bursts over the year-long timescale. This suggests that the evolution of the FRS should be different from impulsive radio flares associated with SNe or GRBs. More similar to the radio lobes of radio galaxies, after a long-term evolution and energy injection, the FRS will likely stabilize and become a PRS. According to this scenario, other more steady PRSs \citep{2017Natur.541...58C,2022Nature_Niu,2024Natur.632.1014B,2024ApJ...976..199I} are likely associated with older FRB engines. Such a speculation is supported by the following fact: In the theoretically-motivated $\rm L_{PRS}-RM$ correlation plot \citep{2020ApJ...895....7Y,2022ApJ...928L..16Y,2024Natur.632.1014B}, the \target~FRS is more deviated from the central line and closer to the upper boundary of the range (Figure \ref{fig:spec_L-RM}). We speculate that as time evolves, it will fall back and get closer to the central line, more consistent with other PRSs if the engine can maintain enough energy injection to the remnant.  

\clearpage

%%%%%%%%%%%Figures in Mian texts%%%%%%%%%%%%%

\begin{figure}[H]
 \centering
 \includegraphics[width=0.85\textwidth,angle=0]{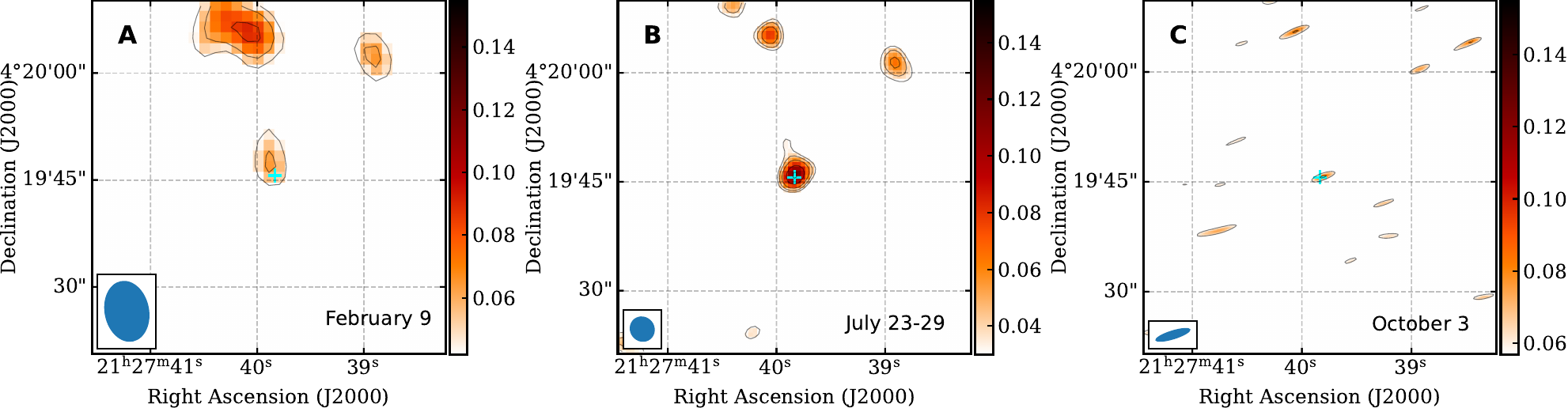}
 %\vspace{-2.0cm}
 \caption{
 \footnotesize \textbf{MeerKAT and VLA sky images of the FRS.} \textbf{A}. MeerKAT 1.5 GHz image in the 2024 February observation. \textbf{B}. The VLA deep image in 2024 late July observations. \textbf{C}. The VLA image in the 2024 October observations. The contours in all images correspond to (1, $\sqrt{2}$, 2, 3, 4, 5) times the 3$\times$rms noises (14, 10 and 19 $\mu$Jy), respectively. The lowest values shown in the images correspond to 3$\times$rms noises. The position of the FRB obtained with EVN \citep{2024ATel16542....1S} is shown as a cyan cross. The synthesized beam is shown as a filled blue ellipse in the bottom left of all panels. The values in the color bars on the right side are in units of mJy. }
 \label{fig:skyim}
\end{figure}

\clearpage

\begin{figure}[H]
    \centering
    \subfloat{
        \includegraphics[width=0.85\textwidth]{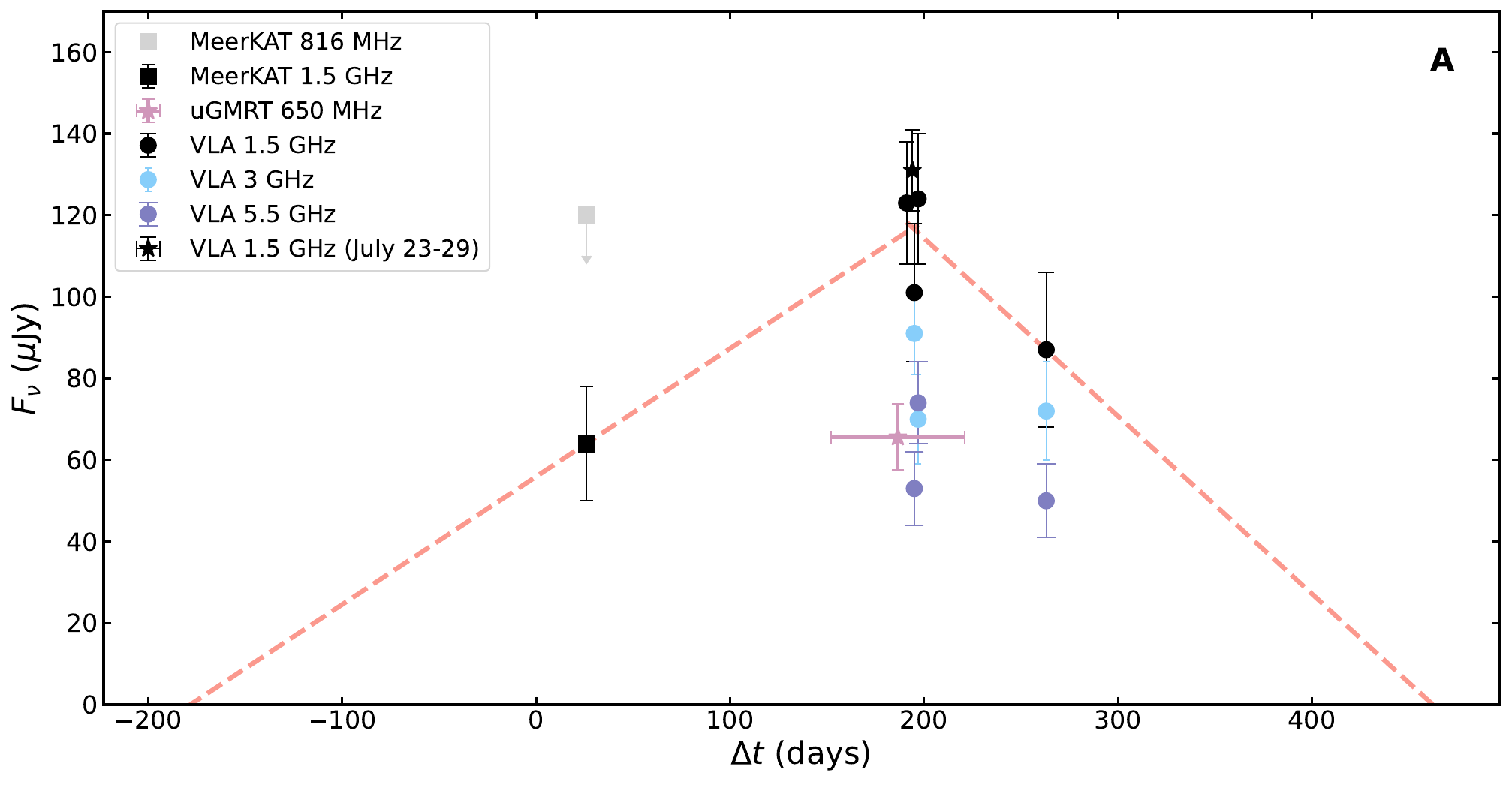}
    }\\
    \subfloat{
        \includegraphics[width=0.85\textwidth]{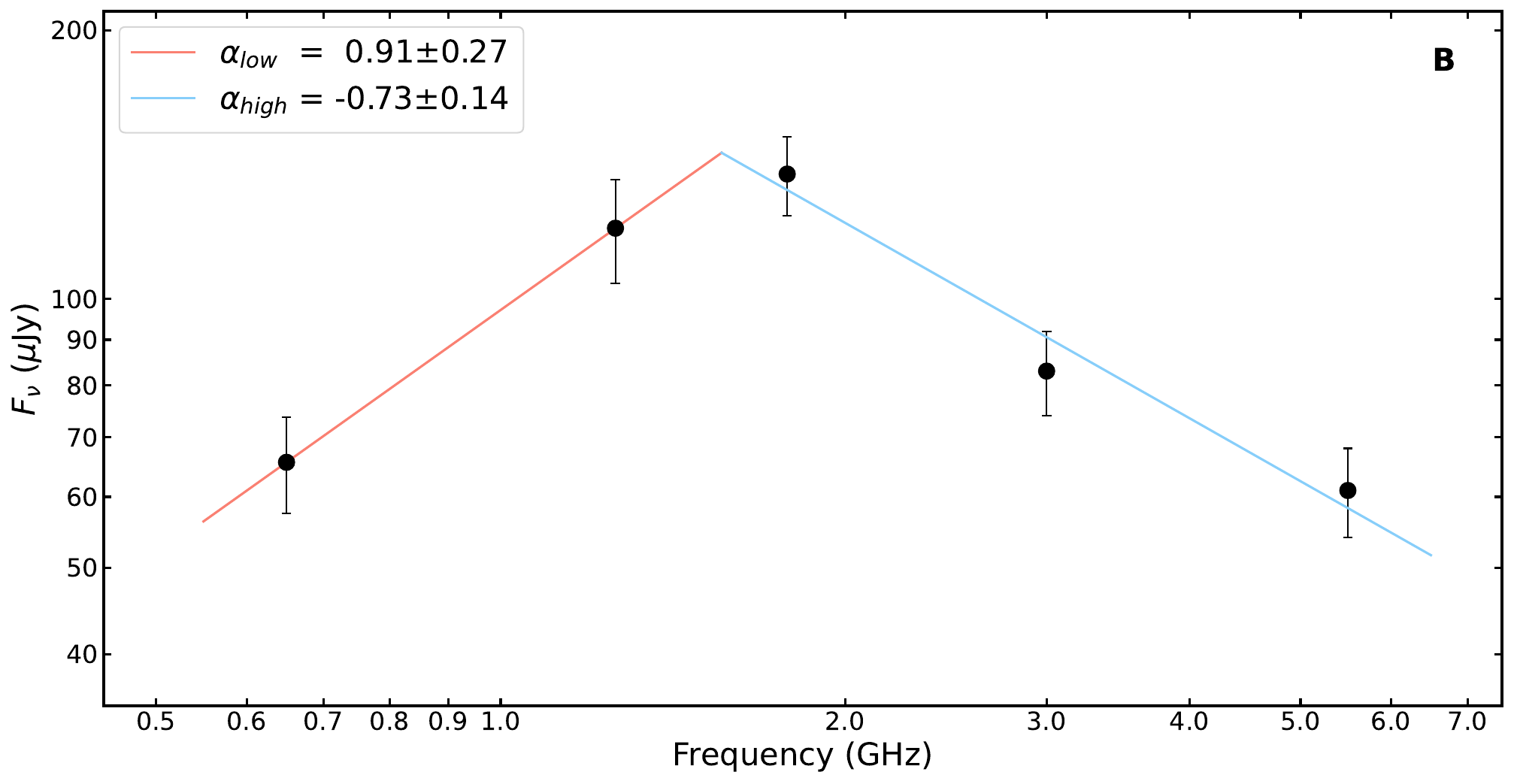}
    }
    \caption{
    \footnotesize \textbf{Multi-band radio light curves of all observations and the broad-band radio spectrum taken in 2024 July.} \textbf{A}. Radio flux densities of the FRS as measured by MeerKAT (UHF- and L-bands) and VLA (L-, S- and C-bands), the flux density measured by uGMRT at a central frequency of 650 MHz \citep{2024ATel16820....1B,2024arXiv241213121B}, and the averaged flux density of the FRS obtained from the 1.5 GHz VLA deep image in the late July observations are shown. The x-axis shows the time elapsed in days since the discovery of \target. The two red dashed lines show the linear fit to the first four and the last four flux density measurements at 1.5 GHz, respectively, with the slope of 0.31$\pm$0.10 and -0.44$\pm$0.31. \textbf{B}. The broad-band radio spectrum of the FRS around 2024 late July. The flux density at 650 MHz is taken from the uGMRT measurement \citep{2024ATel16820....1B,2024arXiv241213121B}. We fit the spectrum with a broken power-law model, which determined the spectral break at 1.56$\pm$0.20 GHz. }
    \label{fig:lc_spec}
\end{figure}

\begin{figure}[H]
 \centering
 \includegraphics[height=7.50in]{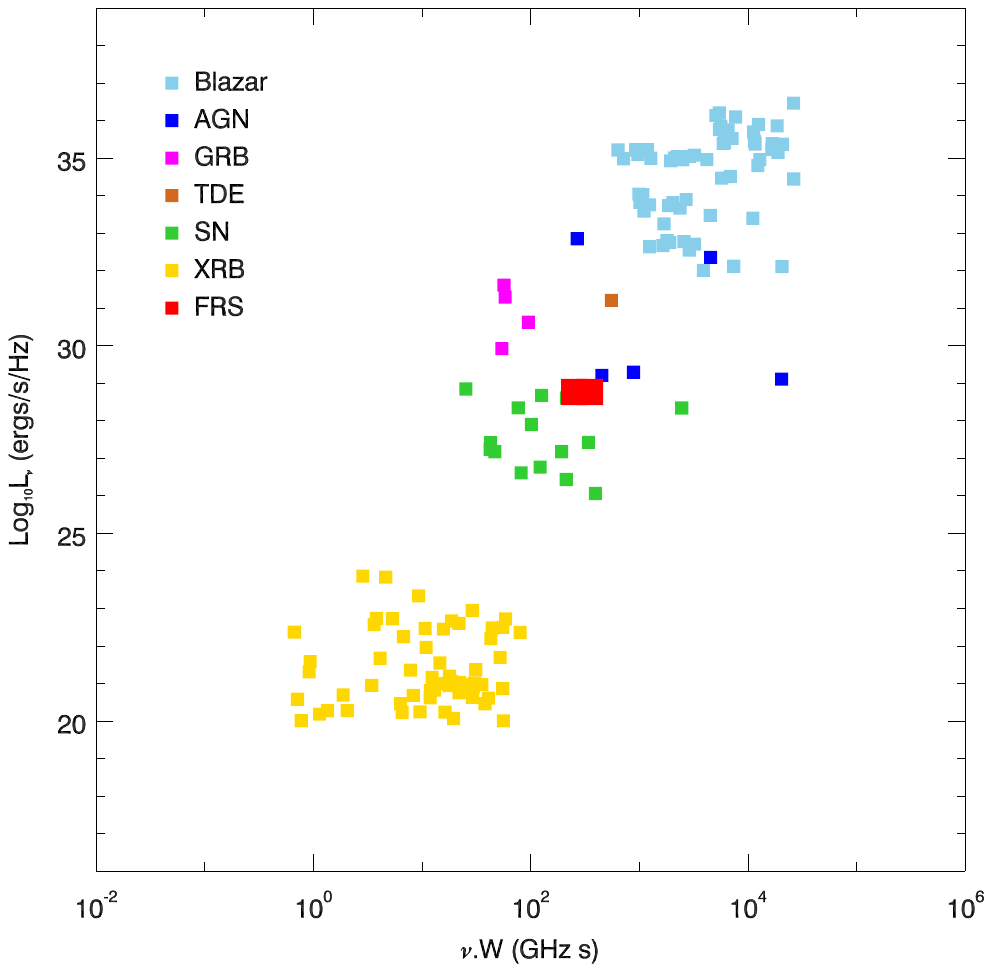}
 \vspace{-2.0cm}
 \caption{
 \footnotesize \textbf{The spectral luminosity for the FRS and several different types of radio transients as a function of their width (${W}$) and frequency ($\nu$).} The data for GRBs, AGN/Blazars, XRBs, TDEs, and SNe are from previous samples \citep{2015MNRAS.446.3687P,2024NatAs...8.1159C}. ${W}$ corresponds to the e-folding time scale in seconds, and $\nu$ corresponds to the frequency in GHz for the flare peak. The two red squares, which overlap, represent the measurements of the e-folding rise and the e-folding decay time scales of the FRS. The AGN sample nearest to the FRS is NGC 7213, a low-luminosity active galactic nucleus (LLAGN); The SNe sample nearest to the FRS is SN 1998bw, see also Figure \ref{fig:FRS_sne}.}
 \label{fig:comparison_transients_w}
\end{figure}

\begin{figure}[H]
 \centering
 \includegraphics[height=7.0in]{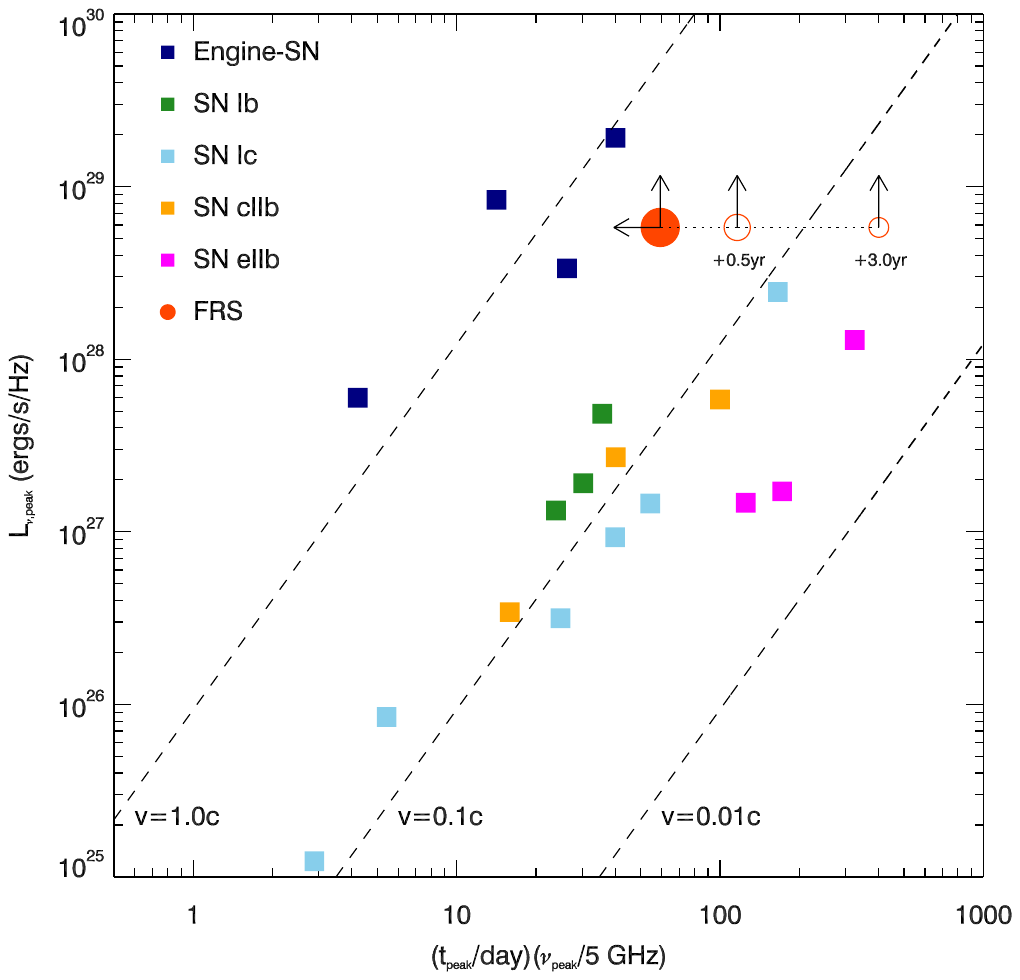}
\vspace{-2.0cm}
 \caption{
 \footnotesize \textbf{Comparison of the radio properties between the FRS and tho radio supernovae.} The FRS and the SNe \citep{2012ApJ...752...78S} are marked as a filled circle (in red) and squares, respectively. The spectral radio luminosity at the flare peak ${L}_{\nu}$ of the FRS is taken as the measurement at 1.5 GHz in late July, and the ${t}_{\rm peak}$ is calculated as the time interval between the FRB discovery and late July (red solid circle). If the FRS is indeed a newly born source, the former corresponds to the lower limit of the peak luminosity and the latter corresponds to the upper limit of the peak time, respectively, and the FRS is consistent with those engine-powered SNe (the GRB-SNe) under the SN shock model \citep{1998ApJ...499..810C}, of which the average velocities of the blast-waves (spectral index $\alpha$=2.5) are over-plotted as dashed lines. The closest samples are then SN 2003lw and SN 2009bb. If the FRS had been born for half a year or more, its peak luminosity would be higher than those SNe Ib/c and SNe II samples with slower blast-wave velocities, and the closest sample is SN 2003L. The red open circles correspond to the cases in which FRS was 0.5 years or 3.0 years old when \target~was discovered by CHIME. 
 }
 \label{fig:FRS_sne}
\end{figure}

%% Also note that the akcnowlodgment environment does not support long amounts of text. If you have a lot of people and institutions to acknowledge, do not use this command. Instead, create a new \section{Acknowledgments}.
\begin{acknowledgments}
We thank the staff at National Radio Astronomical Observatory for scheduling and performing the VLA DDT observations and the staff at South Africa Radio Astronomical Observatory for performing the MeerKAT observations and making the data available. The MeerKAT telescope is operated by the South African Radio Astronomy Observatory, which is a facility of the National Research Foundation, an agency of the Department of Science and Innovation. W.Y. would like to thank Dr. M. Caleb of University of Sydney for providing data of radio transient samples used for our plots. W.Y., X.Z. and Z.Y. acknowledge support from the Natural Science Foundation of China (No.12373050,U1838203,12373049). The \textsc{ztfquery} code was funded by the European Research Council (ERC) under the European Union's Horizon 2020 research and innovation programme (grant agreement n°759194 - USNAC, PI: Rigault). X.Z. analyzed MeerKAT and VLA data; W.Y. contributed to the data analysis; X.Z. and W.Y. proposed the VLA DDT observations and coordinated the multi-wavelength efforts; X.Z., W.Y., and B.Z. contributed to the calculations and the cross-checks; W.Y., B.Z., and X.Z. proposed the interpretations; Z.Y., W.Y., and Y.X. performed optical/infrared, X-ray, and gamma-ray data analysis; X.Z., W.Y., and B.Z. contributed to the primary draft; all authors contributed to the writing and discussion of the paper. MeerKAT data are directly available from the data archive provided by the South Africa Radio Astronomical Observatory. VLA DDT data can be fully accessed from the data archive provided by the National Radio Astronomical Observatory once the proprietary periods expire. $Swift$ data is available from the $Swift$ archive in HEASARC. The ZTF data is available in the NASA/IPAC Infrared Science Archive. All the software packages used in the analysis are publicly available. 
\end{acknowledgments}


%\vspace{5mm}
%\facilities{HST(STIS), Swift(XRT and UVOT), AAVSO, CTIO:1.3m,


%\software{astropy \citep{2013A&A...558A..33A,2018AJ....156..123A},  
%          Cloudy \citep{2013RMxAA..49..137F}, 
%          Source Extractor \citep{1996A&AS..117..393B}
%          }

\newpage
\appendix

\section{Radio Observations and Data Reduction}
\label{sec:Radio_obs} 
\subsection{MeerKAT observations and data reduction}
\label{sec:meerkat_datareduction}
After the discovery of the repeating FRB 20240114A on 2024 January 14, two bursts were detected and reported at L-band \citep{2024ATel16446....1T} in the MeerKAT DDT (PI: Jun Tian) observations performed on 2024 February 9. More burst detections have been reported \citep{2024MNRAS.533.3174T} -- 44 in the UHF-band and 18 in the L-band over the full observing time. The observations consist of an L-band observation and an UHF-band observation, both have a total exposure time of $\sim$1.4 hours and an on-source time of $\sim$1 hour. The visibilities of 2 seconds sampling time were saved for both observations. The L-band data was taken at a central frequency of 1.28 GHz with 4096 channels in total, reaching a bandwidth of 856 MHz; the UHF-band data was taken at a central frequency of 816 MHz with 4096 channels, reaching a bandwidth of 544 MHz. J1939$-$6342, the absolute flux and bandpass calibrator, was observed for about 10 min at the beginning of both observations. In both observations, J2130$+$0502 was observed as the phase calibrator for about 10 min after the on-source time of about 1 hour spent on the field of the target source \target; however, in the UHF-band observation, there is an additional scan with a 2-min exposure on J2130$+$0502 before the scan on the target field. Data reduction (flagging, calibration, imaging) was done using the dedicated, semi-automatic pipeline \textsc{oxkat} \citep{2020ascl.soft09003H}. We executed the pipeline step-by-step for both observations and ceased it after the direction-independent self-calibrations were completed (direction-dependent self-calibrations were not performed): These firstly involved an initial examining, basic flagging, and calibration (delay, bandpass and gain) of data through Common Astronomy Software Application package (hereafter \textsc{casa}; \citealt{2007ASPC_McMullin,2022PASP..134k4501C}); Afterwards, a further flagging process through the package \textsc{tricolour} \citep{2022ASPC..532..541H} was performed for the target field before imaging with \textsc{wsclean} \citep{2014MNRAS.444..606O}; %and producing a mask for further calibration. 
Finally, we direction-independently self-calibrated the data with the package \textsc{cubical} \citep{2018MNRAS.478.2399K}. We did a further flagging for the calibrated target data using \textsc{rflag} in \textsc{casa} to edit bad data that still remained. We finally imaged the data in total intensity (Stokes I) using \textsc{tclean} in \textsc{casa}. We used a Briggs weighting scheme with a robust parameter of $0.0$ to achieve compromise between the side-lobe effects across the field and sensitivity. Besides, the auto-thresholding algorithm was invoked for the clean. Images were primary-beam corrected with \textsc{katbeam}\footnote{\url{https://github.com/ludwigschwardt/katbeam}}. We made use of the \textsc{casa} task \textsc{imfit} to measure the flux density of the source for all epochs by fitting a Gaussian component in the image plane. The local root-mean-square (rms) noise in each observation was obtained from a close-by source-free region. A summary of the MeerKAT observations is listed in Table~\ref{tab:meerkat+vla_observations}.
 
\subsection{VLA observations and data reduction}
\label{sec:vla_datareduction}
The Karl G. Jansky Very Large Array (VLA) DDT multi-band observations (Project ID: 24A-488 and 24B-467; PI: Xian Zhang) were performed in late July 2024 and October 3, allowing us to investigate the temporal and spectral properties, and the compactness and the radio position of the variable radio continuum counterpart, i.e., the FRS associated with \target. The three epochs of observation of the project 24A-488 were carried out within one week (July 23--29), consisting of one epoch, single L-band observation taken on July 23 and two epochs of multi-band observation (L-, S-, C-band) taken on July 27 and 29 respectively. The central frequencies of the three observing bands are 1.5 GHz, 3 GHz, and 5.5 GHz with observing bandwidths of 1 GHz, 2 GHz, and 2 GHz, respectively. Data integration times are all 3 seconds for all observations. The telescope was in its B configuration during this observing project; Project 24B-467 was executed once on October 3. We also include this observation in this report. The observing central frequencies and the corresponding bandwidths are the same as those of the first observation project (24A-488). The data integration time is 2 seconds and the telescope configuration is BnA. For both projects, the calibrators are the bandpass and absolute flux calibrator 3C 286 (J1331$+$305) and the phase calibrator J2130$+$0502. Data reduction (i.e., data flagging, calibration) was done using \textsc{casa}. We performed further flagging on the target field and then subsequently imaged its Stokes I data using the deconvolution algorithm \textsc{tclean}. To balance sensitivity while suppressing side-lobes effects, we imaged with a Briggs weighing scheme (robust=0). Images were primary beam corrected with \textsc{widebandpbcor}. We made use of the \textsc{casa} task \textsc{imfit} to measure source flux densities by fitting a Gaussian model in the image plane. A summary of the VLA observations is listed in Table~\ref{tab:meerkat+vla_observations}.

\section{Radio imaging analysis and source detections}
\begin{figure*}
\centering
\includegraphics[width=0.85\textwidth,angle=0]{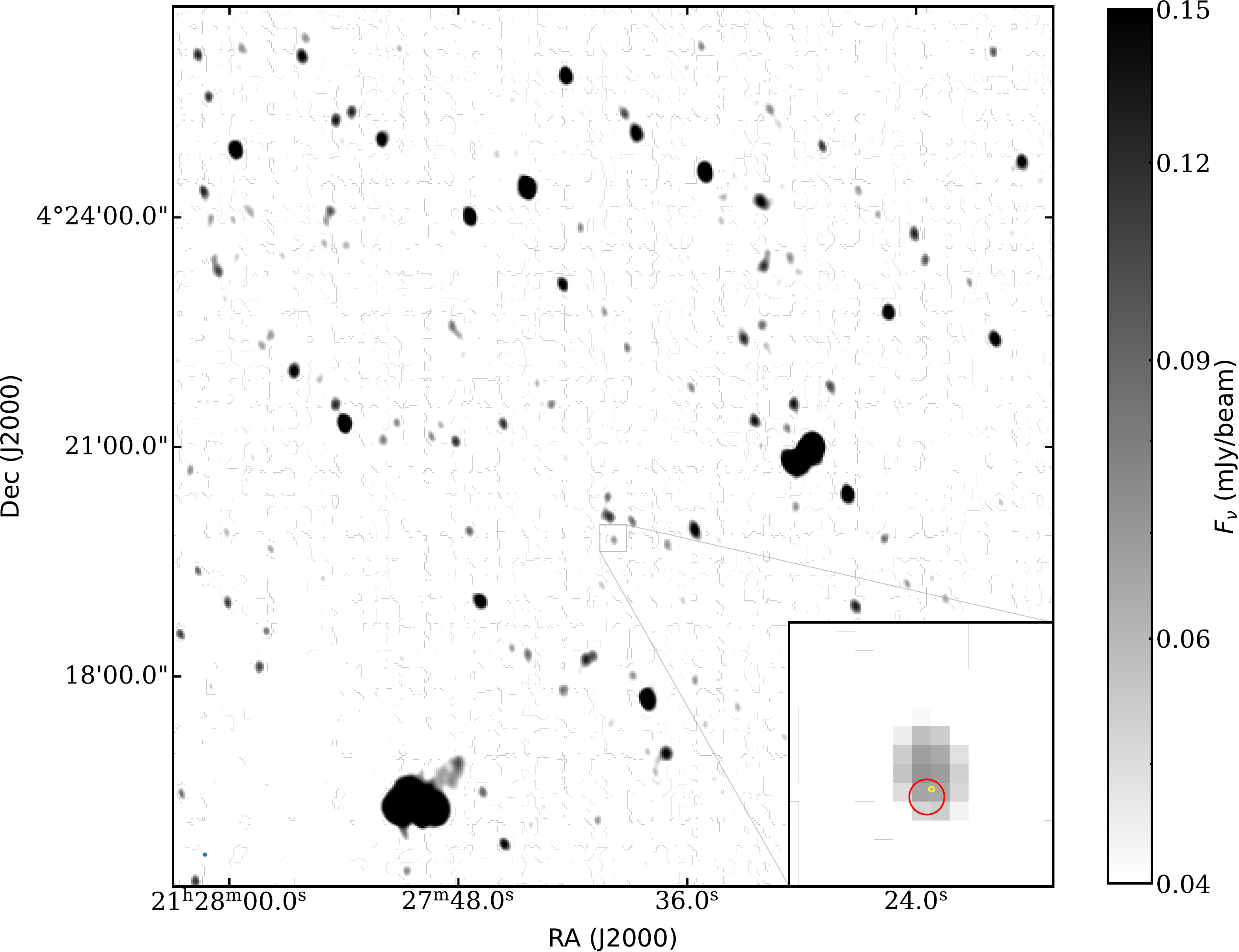}

\caption{
\footnotesize \textbf{MeerKAT sky image of \target~and the FRS seen at 1.28 GHz.} Image of the field of \target~and the FRS at 1.28 GHz observed with MeerKAT in 2024 February. The position of the brightest burst detected by MeerKAT \citep{2024MNRAS.533.3174T} is shown with the red circle. The position of the FRB obtained with EVN \citep{2024ATel16542....1S} is shown with the yellow circle. The lowest value shown in the image is 3$\times$rms noise.}
\label{fig:image_meerkat}
\end{figure*}

\begin{figure*}
\centering
\includegraphics[width=0.85\textwidth,angle=0]{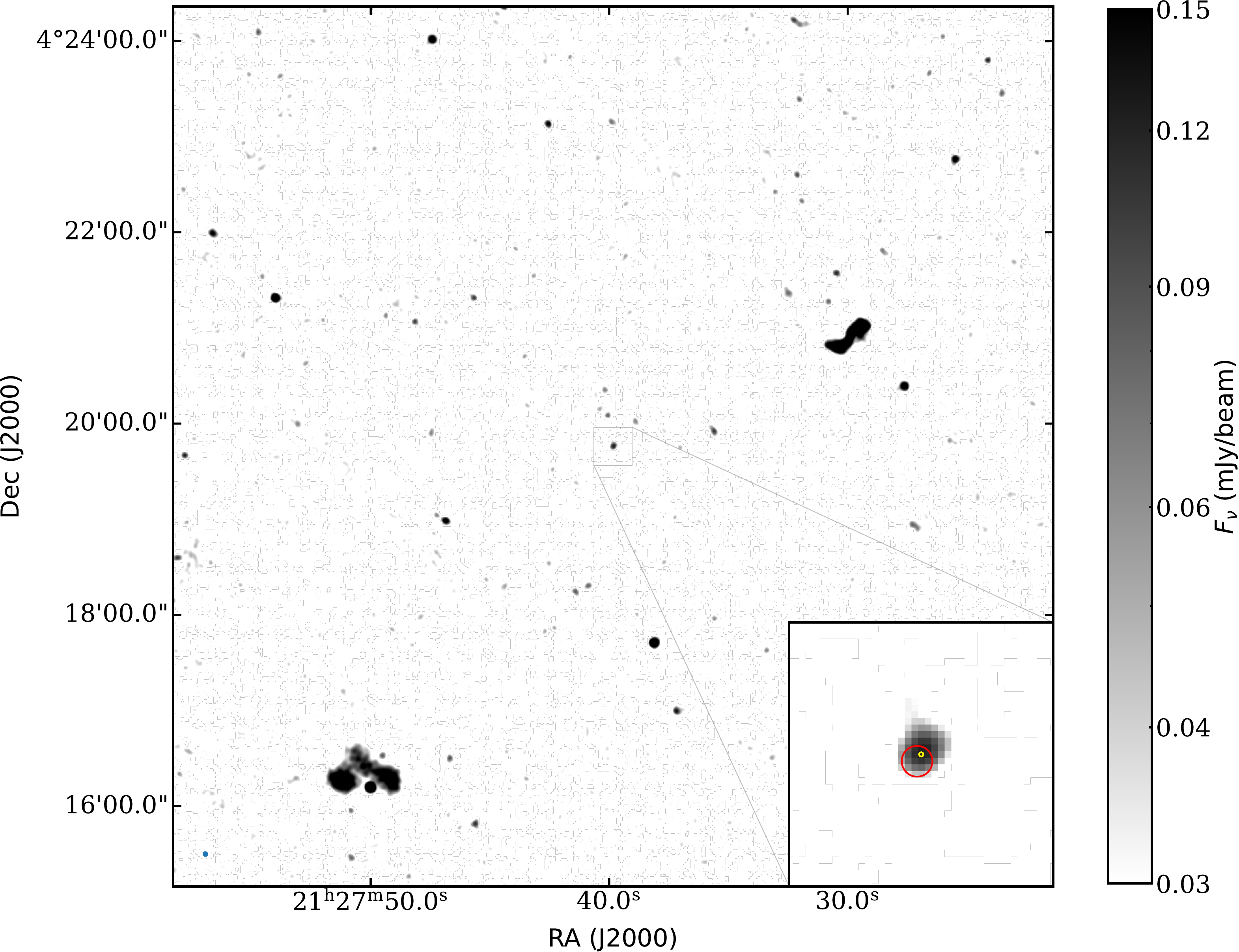}

\caption{
\footnotesize \textbf{VLA deep sky image of \target~and the FRS seen at 1.5 GHz.} The deep image of the field of \target~and the FRS at 1.5 GHz obtained with the VLA in 2024 July. The position of the brightest burst detected by MeerKAT \citep{2024MNRAS.533.3174T} is shown with the red circle. The position of the FRB obtained with EVN \citep{2024ATel16542....1S} is shown with the yellow circle. The lowest value shown in the image is 3$\times$rms noise.}
\label{fig:deep_image_vla}
\end{figure*}

In the MeerKAT L-band observation taken on 2024 February 9, we discovered the FRS at the position of (R.A., Decl.)~[J2000] = (21h27m39.82s, 4\arcdeg19\arcmin47.106\arcsec) with 1 $\sigma$ statistical uncertainty of (0.65\arcsec, 0.89\arcsec), which is spatially coincident with the FRB positions given by MeerKAT burst localization \citep{2024ATel16446....1T,2024MNRAS.533.3174T} and EVN preliminary burst localization \citep{2024ATel16542....1S} (Figure \ref{fig:image_meerkat}). %This position motivated us to try to more accurately constrain its position with VLA. 
The VLA DDT observation campaign (24A-488 and 24B-467) has led to four L-band epochs, three S-band epochs, and three C-band epochs of observations. We show the deep VLA sky image of the FRS field at 1.5 GHz corresponding to observations in late 2024 July in Figure \ref{fig:deep_image_vla}, during which the FRS was most significantly detected, with a flux density of 131$\pm$10 $\mu$Jy. The FRS remained unresolved in all epochs of observations in different observing bands. To better constrain the size in the image plane and also the radio position of the FRS, we stack observations in C-band in the uv-plane and then imaged the Stokes I intensity. Fitting a Gaussian model in the image plane for the FRS emission component, the source is found unresolved with an upper limit of the size of 1.4\arcsec$\times$0.5\arcsec. The radio source position is constrained to (R.A., Decl.)~[J2000] = (21h27m39.84s, 4\arcdeg19\arcmin45.598\arcsec) with 1 $\sigma$ statistical uncertainty of (0.06\arcsec, 0.04\arcsec), which is spatially coincident with the FRB positions given by MeerKAT burst localization \citep{2024ATel16446....1T,2024MNRAS.533.3174T} and EVN preliminary burst localization \citep{2024ATel16542....1S}.

\section{Radio variability analysis and the radio light curve}
\begin{figure}[H]
 \centering
 \includegraphics[height=7.0in]{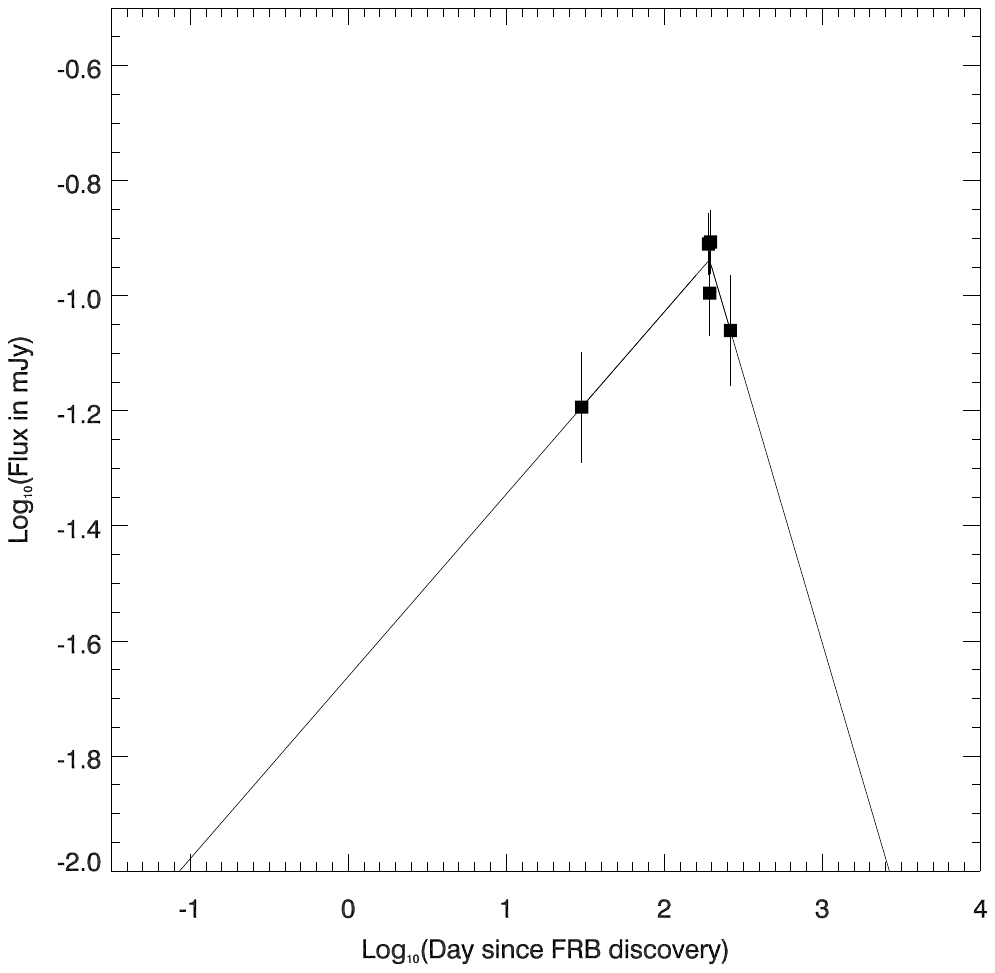}
 \vspace{-1.0cm}
 \caption{
 \footnotesize \textbf{ The 1.5 GHz light curve measured with MeerKAT and VLA observations.} Logarithmic model fit to the rise (four measurements) and the possible decay (four measurements) is over-plotted. A logarithmic rise would indicate a rise started much later than that of a linear rise shown in Figure \ref{fig:lc_spec}. The possible decay, if true, would still need at least a few years or more.}
 \label{fig:lc_loglog}
\end{figure}

In Table \ref{tab:meerkat+vla_observations}, Figure \ref{fig:lc_spec}, and Figure \ref{fig:lc_loglog}, we show the flux density measurements and the corresponding light curves for the FRS, respectively. In the L-band, the FRS rose in flux density by a factor of 2 between the MeerKAT observations in February and the VLA observations in late July; then showed a likely decline in flux density between late July and early October. The three VLA 2024 July observations were performed within a week, confirming the rise with consistent flux detections. Such a flux rise by 100\% has never been seen before in any PRSs, which defines the radio continuum counterpart as a unique flaring radio source (see below and also Figure \ref{fig:flux-flux}). However, due to the sparse coverage of the sensitive radio observations in the early rise (probably less than 6 months), the radio flare could have been brighter. With these flux density measurements, we were able to measure the rise and the decay time scales, bearing in mind that the flare peak was probably missed such that the rise time scale could be shorter and the decay time scale could be longer. Therefore, an average of the rise and the decay time scales should be a better indicator of the characteristic variability time scale of the FRS.  

We also characterize the variability of the radio continuum counterpart associated with \target~with two parameters -- modulation index ($m_{\rm o}'$) and variability significance ($\chi^2$), as detailed in previous investigations of the variability of radio transients/variables \citep{2021ApJ_Sarbadhicary,2023MNRAS_meerkat,2023ApJ...959...89Z}. We found the FRS did not vary on the shorter, intra-week time scale in late July, as evidenced by the flux density measurements (within uncertainties). Therefore, we were able to robustly measure the flux density of the FRS in late July by stacking three VLA observations taken on July 23, 27, and 29. This helped characterize the apparent rising phase (2024 February to July) and the apparent declining phase (2024 July to October) in the 1.5 GHz light curve. During the apparent rise between February and late July, the modulation index and variability significance of the radio source are 48.6$\pm$14.4\% and 15.2 at 1.5~GHz, suggesting that the source increased in flux density significantly. Similarly, during the apparent decline phase between late July and October, the modulation index and variability significance of the radio source are 28.5$\pm$15.7\% and 4.2 at 1.5~GHz. In the following, we demonstrate that the variability seen at 1.5 GHz is intrinsic to the FRS, rather than potential systematic effects due to inappropriate calibration or scintillation.

%The modulation index and variability significance of the radio source are 34.9$\pm$12.0\% and 15.2 at 1.5~GHz during the observation campaign, suggesting that the source appeared to vary from 2024 February to October, while potential variability at 3~GHz and 5.5~GHz was not significant, as shown in Table~\ref{tab:meerkat+vla_observations} and Figure \ref{fig:lc_spec}. In addition, we found the FRS did not vary on the weekly time scale at 1.5 GHz in late July, as evidenced by the flux density measurements (within uncertainties) in late July. Therefore, we were able to robustly measure a flux density in late July of the FRS by stacking three VLA observations taken on July 23, 27, and 29 and identify two phases in the 1.5 GHz light curve -- an apparent rising phase (2024 February to July) and an apparent declining phase (2024 July to October): During the apparent rise between February and late July, the modulation index and variability significance of the radio source are 48.6$\pm$14.4\% and 15.2 at 1.5~GHz, suggesting that the source increased in flux density significantly. Similarly, during the apparent decline phase between late July and October, the modulation index and variability significance of the radio source are 28.5$\pm$15.7\% and 4.2 at 1.5~GHz. In the following, we show that the variability seen at 1.5 GHz is intrinsic to the FRS, rather than possible systematic effects due to inappropriate calibration or scintillation.

\subsection{Flux measurements across telescopes and observations}
We first excluded the possibility that the variability detected from the FRS at 1.5 GHz was due to calibration effects, by investigating the variability properties of sources in the same field-of-view. We investigated the variability of the field sources at 1.5 GHz in the period from 2024 February to late July. We ran the \textsc{pybdsf}\footnote{\url{https://www.astron.nl/citt/pybdsf/index.html}} package to extract radio sources from the MeerKAT image and VLA deep images obtained with observations in 2024 February and late July. These radio point sources were identified with the following criteria:
\begin{itemize}
\item The peak intensity (Jy/beam) should be 0.9 times higher than its integrated flux (Jy) for the 1.5 GHz image, intended to select point sources.
\item The S/N (peak intensity / local rms noise) should be greater than 5, intended to select sources with high significance. 
\item The peak intensity is higher than that of the FRS in late July, intended to select brighter sources. 
\end{itemize}

In total, we detected 397 sources in the MeerKAT image and 58 sources in the VLA deep image. These extracted sources were also visually checked to make sure that they correspond tp point-like sources. We then cross-matched the detected point sources between the MeerKAT and VLA images. 
%Comparing with the detected `point-like' sources in the MeerKAT image, 
We cross-matched 17 nearby point-like sources within 10 arcmin of the phase center -- the position of the FRS in the VLA deep image. Out of these sources, only 1 of them varied with a reduced $\chi^2$ as large as 5, but still less varied than the FRS, suggesting no variability can be produced by unknown systematic calibration effects between the two epochs and robust detection of the intrinsic variability of the FRS from February to late July. 

\begin{figure}[H]
 \centering
 \includegraphics[height=2.7in]{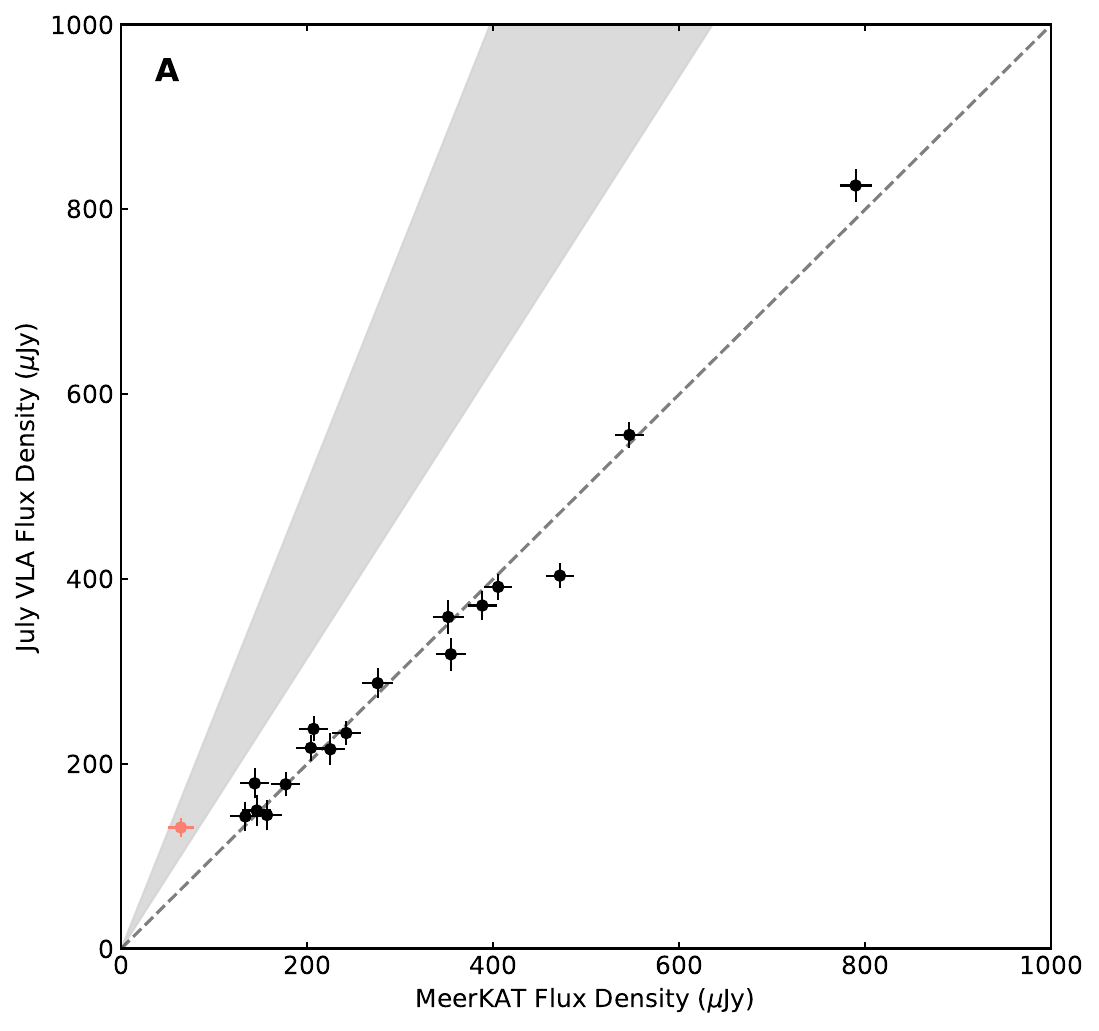}
 \includegraphics[height=2.7in]{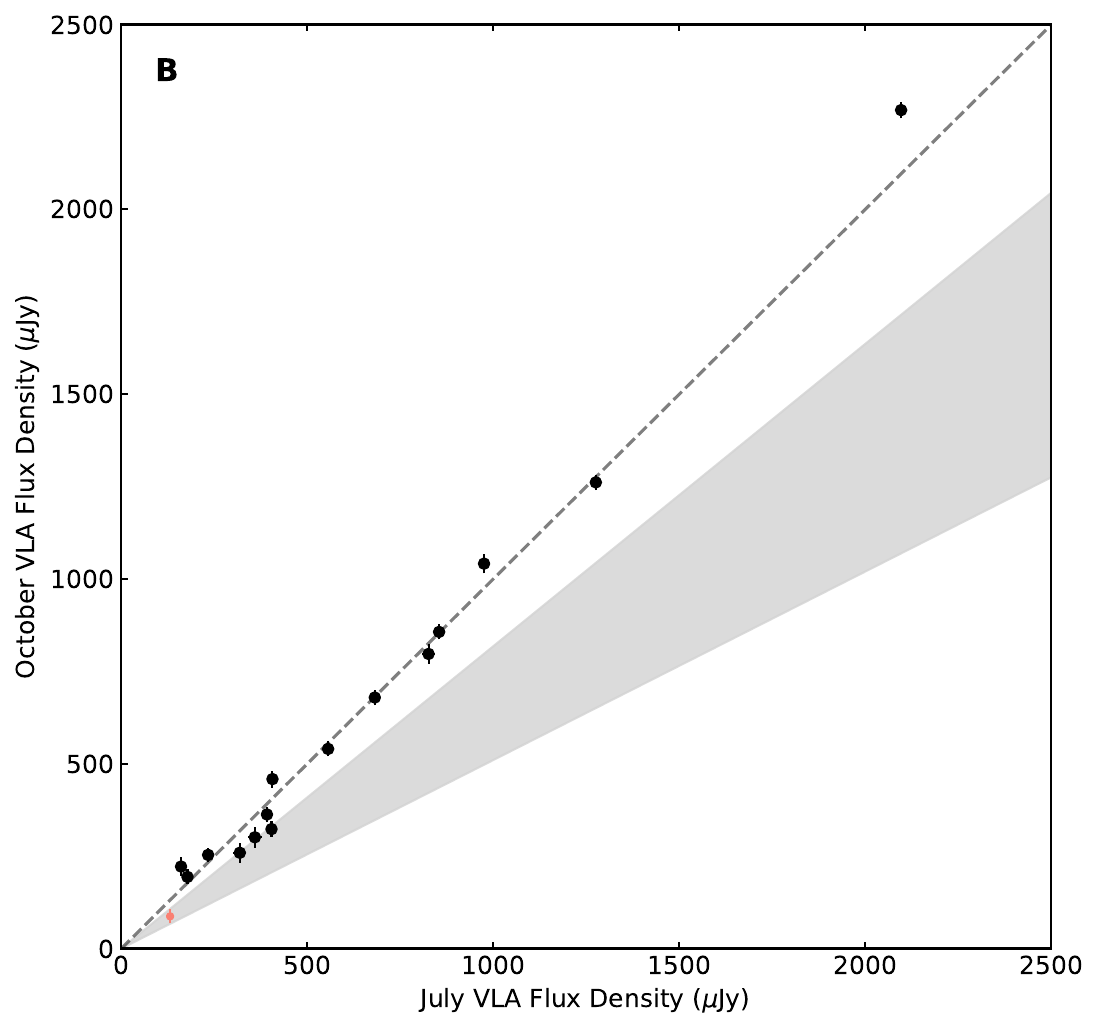}
 \caption{\footnotesize \textbf{Flux densities of the cross-matched point sources between observations.} \textbf{A}. Flux densities of the cross-matched 17 point-like sources and the FRS measured with the VLA observations taken in late July versus those measured in the MeerKAT observations taken in February, respectively. The dashed line represents the relation of a slope of 1, indicating cases when the flux density measured with MeerKAT in February is equal to that measured with the VLA in July for the same source. The regime shaded in gray represents the $1\sigma$ range of the slope corresponding to the rise in flux density. The slope for the FRS was used to estimate the e-folding rise time. \textbf{B}. Flux densities of the cross-matched 15 point-like sources and the FRS measured with the VLA observations taken in October versus those taken in July, respectively. The scheme is similar to \textbf{A}. }
 \label{fig:flux-flux}
\end{figure}

We determined that the VLA flux density measured in July is $A$ = 2.05$\pm$0.47 times the MeerKAT flux density for the FRS (Figure \ref{fig:flux-flux}), yielding an e-folding rise time of $\tau_{r}$ = 235$\pm$76 days (the time for the radio flux density to increase by a factor of $A$ was 168.33 days). Based on light-crossing time argument, the e-folding rise time scale $\tau_{r}$ gives the upper limit on the radius as ${\rm R}\leq c\tau_{r}\delta(1+z)^{-1}$ (where $\delta$ is the Doppler factor and $z$ the redshift 0.13), only $\sim0.18\delta$ pc, significantly smaller than the best constraint ${\rm R\leq0.35}$ pc obtained for the PRS of FRB~20121102A with Very Long Baseline Interferometry \citep{2017Natur.541...58C}. 

Similarly, we also compared the measured flux density of the field sources between VLA observations taken in October and July. We cross-matched 15 point-like sources within 10 arcmin of the position of the FRS. Only 2 of the cross-matched sources varied with reduced $\chi^2$ larger than 5. The FRS was moderately varied during this period. The VLA flux density measured in July is $A$ = 1.51$\pm$0.35 times that measured in October for the FRS, therefore we determined the e-folding decay time as  $\tau_{d}$ = 168$\pm$95 days (the time for the radio flux density to decrease by a factor of $A$ is 68.67 days). Both $\tau_{r}$ and $\tau_{d}$ are used in the plot of Figures \ref{fig:comparison_transients_w} and \ref{fig:comparison_transients_efolding}. 

\subsection{Investigation of possible contribution from scintillation}
Compact radio sources are known to show flux variation due to scintillation, particularly as a consequence of the small-scale inhomogeneities in the ionized component of the interstellar medium (ISM). In the following paragraph, we discuss the predicted scintillation properties of the FRS as inferred with our measurements of the variability significance and flux variation coefficient based on the formalism listed in \citep{1998MNRAS.294..307W}, similar to previous studies \citep{2023ApJ...959...89Z}.

Using the description of \citet{1992_Narayan}, the scintillation properties have been parameterized in \citet{1998MNRAS.294..307W} as ``scattering strength": $\xi = ({\nu_0}/{\nu})^{17/10}$, $\nu$~is the observing frequency and $\nu_0$~is the transitional frequency which we estimate as $\nu_0$ = 10.8~GHz based on pyne2001\footnote{\url{https://pypi.org/project/pyne2001/}}, a python wrapper around the original FORTRAN implementation of the NE2001 Galactic free electron density model \citep{2002_Cordes1,2003_Cordes2}. $\xi$ = 1 corresponds to the critical value at which the ISM inhomogeneities introduce a substantial phase change (about half a radian) across the first Fresnel zone, which is a characteristic property of the scintillation screen and has the angular radius of $\theta_{\rm F} = \sqrt{{c}/{2 \pi \nu D}}$ \citep{1992_Narayan}, in which $D$~is the angular distance to the source. The observing frequency at which there was significant variability and flux density increase from 2024 February to July during the campaign came from 1.5 GHz. The type of scintillation we discuss here belongs to the refractive scintillation -- slow and broadband in the strong regime ($\xi\gg$~1, $\nu_0$ $>$~$\nu$), in which the wavefront is highly corrugated on scales smaller than the first Fresnel zone. In the context of refractive scintillation, the modulation index is $m_{\rm p} = \xi^{-1/3} = ({\nu}/{\nu_0}) ^{17/30}$, the angular radius of the scattering disk at frequency $\nu$ is $\theta_{\rm r} = \theta_{\rm F} \xi =\theta_{\rm F,\nu_0} ({\nu_0}/{\nu})^{11/5}$ and the refractive time scale for a compact source with size smaller than the angular size of the scattering disk is $t_{\rm r} \sim 2({\nu_0}/{\nu})^{11/5}$. 

The radio flare (for both rise and decline) seen at 1.5~GHz is not caused by scintillation from a Galactic scattering disk, since: (1) the observed modulation index at 1.5~GHz is around 50\%, which is not consistent with that expected from scintillation, as listed in Table~\ref{ta:predictions}; and (2) the expected variation time scale for the compact radio source is 6.4 days if it's scintillated at 1.5~GHz, much less than the time scale of the rise or the decline of the radio flare. In addition, the source did not show significant flux density variation on the time scale during the VLA observations between July 23 and 29. Therefore, we conclude that the radio variability of the FRS we detected is intrinsic. It is worth noting that the following spectral analysis independently supports the radio variability as intrinsic to the FRS. 

\section{Radio spectral analysis and the spectral break}

We investigated the radio spectra and the spectral  evolution of the FRS for the radio observations taken in February, July, and October (as shown in Figure \ref{fig:spectral_evolution}). We fitted a power-law model to the multi-band data in the form of ($F_{\nu} \propto \nu^{\alpha}$, where $F_{\nu}$ is the observed flux density at the frequency $\nu$, $\alpha$ is the spectral index). The MeerKAT observation in L-band can be split into two sub-bands (each with a bandwidth of 428 MHz) centered at 1.07 GHz and 1.5 GHz, respectively. The two sub-bands were imaged and the flux densities of 92$\pm$17$\mu$Jy at 1.07 GHz and 64$\pm$14 $\mu$Jy and 1.5 GHz were measured, yielding an in-band spectrum index of -1.1$\pm$0.8, consistent with an optically-thin radio emission beyond $\sim$1 GHz. We did not detect the FRS with a 3 $\sigma$ upper limit of 120 $\mu$Jy/beam at 812~MHz (in UHF-band), which might imply that there was a spectral break below $\sim$1 GHz. However, the measured flux densities in sub-bands are consistent with measurement errors and the bandwidth coverage was sparse ($<$1 GHz), so we could not determine the exact type of the spectrum and the frequency below which the radio spectrum turned over. We observed the FRS for three epochs with VLA in 2024 late July. With these observations, neither did we find any significant flux density variations in the three observing bands, nor spectrum variations of high significance -- the measured spectral indices are 0.97$\pm$0.68 (L-band in-band spectrum index), -0.49$\pm$0.17 and -0.42$\pm$0.16 for observations taken on July 23, 27 and 29. Concatenating all visibility data in each band, we obtained deep images for individual bands in late July, resulting in an averaged spectrum index of -0.6$\pm$0.1. Splitting the L-band data into two sub-bands with frequencies centered at 1.26 GHz and 1.78 GHz, and then imaging both,  respectively, we identified a potential spectral break which peaked between 1.26 GHz and 1.78 GHz. Taking into account the flux density measured with uGMRT at lower frequencies (the midpoint of the observations is July 18), we found a prominent spectral peak, which will be modeled and discussed in the following section. On October 3, the FRS was observed with VLA and the broad-band (1--6.5 GHz) spectrum can be described with a power-law model with an index of -0.43$\pm$0.21, indicating the possibility that the spectral break was progressively evolving towards even lower frequencies ($\lesssim$1GHz).

\section{Radio spectral modeling and physical considerations}

Our VLA DDT observations in 2024 late July (within a week) have shown that the radio source was consistent with a stable spectrum on the intra-week time scale. The 4 epoch uGRMT observations between 2024 June 14 and August 22 yielded a detection with a flux density of 65.6$\pm$8.1 $\mu$Jy at 650 MHz \citep{2024ATel16820....1B,2024arXiv241213121B}, corresponding to a detection of the source at the mid-time of 2024 July 18 (MJD 60509), close to the mid-time of our VLA observations in July. Therefore, we were able to model the observed spectrum of the FRS with the flux density measurements obtained from the uGMRT detection and our VLA multi-band detections. 

The piecewise function -- the two power-law functions used to describe the broadband spectrum, takes the form of:

\begin{equation}
F_\nu=
\begin{cases}
a \nu^{\alpha_{\rm low}} &\text{if } \nu < \nu_{\rm b} \\ 
b \nu^{\alpha_{\rm high}} &\text{if } \nu \geq \nu_{\rm b} \\
\end{cases} 
\end{equation}
where $\alpha_{\rm low}$ and $\alpha_{\rm high}$ represent the power-law spectral indices; $a$ and $b$ are normalizations; $\nu$ is the central frequency and $\nu_{\rm b}$ is the fitted break frequency. The spectral break frequency is constrained to $\nu_{\rm b}$=1.56$\pm$0.20~GHz, at which the model flux density is 145.8$\pm$14.5 $\mu$Jy. Since the spectrum in the high-frequency range is well measured, a potential variation of the low-frequency flux density does not cause a significant change in the measurement of the spectral break and the corresponding flux density. For example, if we reduce the uGMRT flux density by half and perform the same spectral fit, we get the break frequency at 1.44$\pm$0.11 GHz and a flux density of 154.9$\pm$14.8 $\mu$Jy, consistent with the above values. So the determinations of the spectral break and peak flux density are robust. 

\begin{figure*}
 \centering
 \includegraphics[width=0.80\textwidth,angle=0]{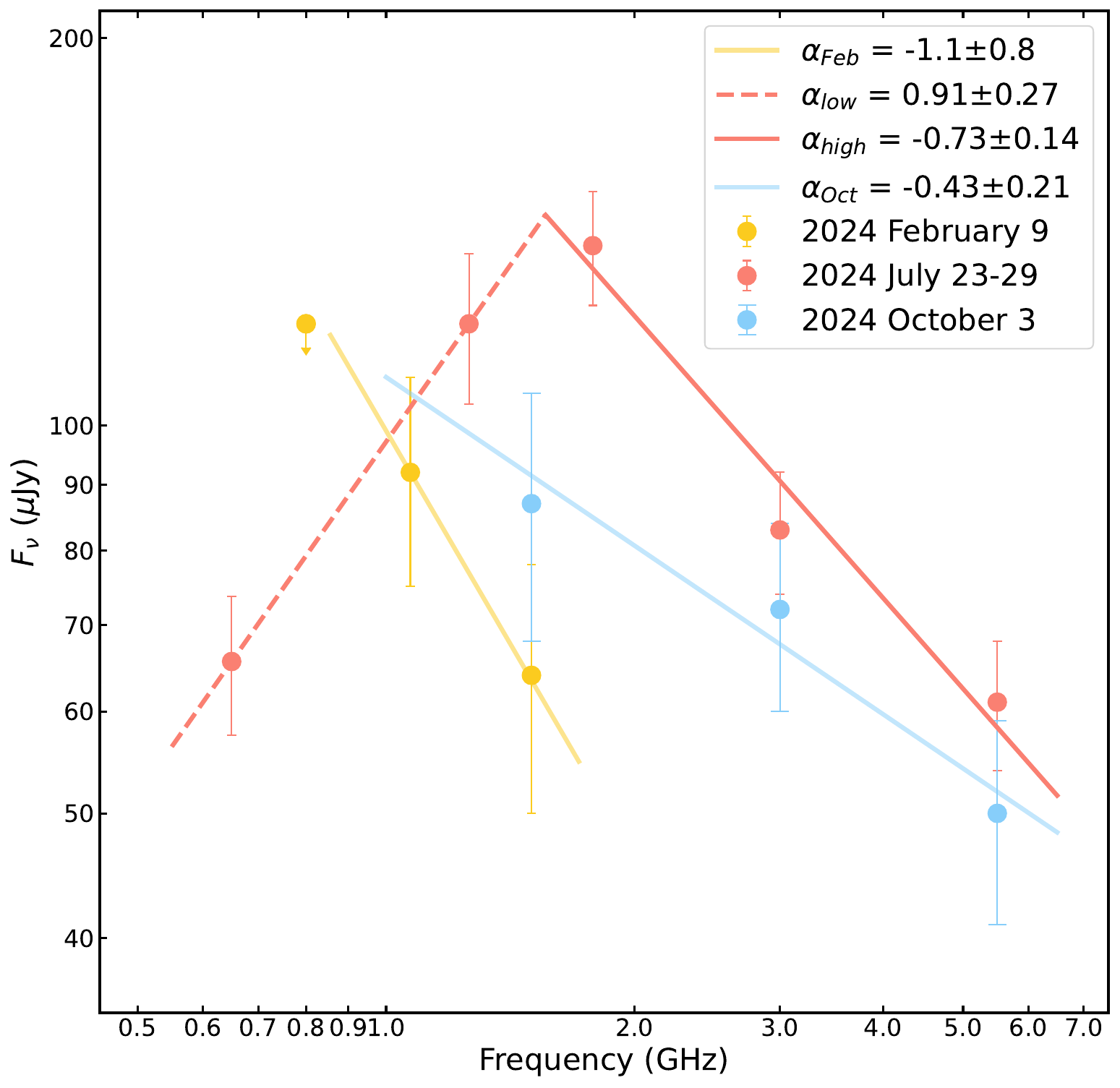}
 \caption{
 \footnotesize \textbf{The broadband radio spectra and the corresponding spectral model fits.} The spectral measurements are obtained from the MeerKAT observation in February (yellow) and the VLA observations in July (red) and October (blue) as well as the measurement made with the uGMRT observations taken from late June to late August 2024 (at 650 MHz). The model fits are over-plotted to cover the entire frequency boundaries of the measurements. The evolution of the spectra is consistent with an increase and a decrease in the spectral break frequency during the rise and the decay of the radio flare, respectively. The evolution of the spectral break frequency under the SSA interpretation implies the plasma of the FRS evolved dynamically, having a varying frequency threshold $\nu_{\tau=1}$ which separates the optically thick and the optically thin regimes. }
 \label{fig:spectral_evolution}
\end{figure*}

In the source rest frame, the spectral break is ${\nu^\prime}_{\rm b} = \nu_{\rm b}(1+z) = 1.76\pm0.23$ GHz ($z=0.13$). In addition, the spectral indices below and beyond the spectral break frequency $\nu_{\rm b}$ are obtained as $-0.91\pm0.27$ and $0.73\pm0.14$, respectively. The modeled spectral indices, spectral break frequency, and corresponding flux density are also shown in Figure \ref{fig:lc_spec} and Figure \ref{fig:spectral_evolution}.

%In the SSA spectral model, the frequency at which the optical depth of the plasma equals to 1 ($\tau$=1) is $\nu_{\tau=1}$ instead of the peak frequency, i.e., the observed spectral break frequency $\nu_{break}$. For a power-law electron  distribution with an index $p$ ($1-2\alpha_{high} = 2.46$), $\nu_{\tau=1} = 0.79\ \nu_{break} = 1.24\pm0.16$ GHz\cite{1970ranp.book.....P}, while the flux density at $\nu_{\tau=1}$ is modeled as $F_{\nu_{\tau=1}} = F_{\nu_{break}} (\nu_{\tau=1}/\nu_{break})^{-0.73\pm0.14} = 172.4\pm29.1\ \mu$Jy. 

We can exclude the possibility that the spectral break (at which the opacity is under $\tau = 1$ condition) in the broadband spectrum was due to free-free absorption (FFA). The flux density at $\nu_{\tau=1}$ (i.e., ${\nu^\prime}_{\rm b}$) for FFA can be expressed as:

\begin{equation}
F_\nu = {\frac{2kT_B \nu_{\tau=1}^{2}}{c^{2}}}{\frac{\pi R^{2}}{D^{2}}}  
\label{eq:equation_self_obs}
\end{equation}
where $k$ is the Boltzmann's constant, $\pi R^{2}/D^{2}$ represents the solid angle subtended by the emission region with a size radius of $R$ at a luminosity distance of $D$. Rearranging equation~\ref{eq:equation_self_obs} we have $T_{\rm B}=F_{\nu}c^{2}D^{2}/2k\pi\nu_{\tau=1}^{2}R^{2} = 1.04\times10^{7}\ (F_\nu/\mu\rm Jy)(D_{\rm Gpc}/{\rm Gpc})^{2}(R_{\rm pc}/{\rm pc})^{-2}(\nu_{\tau=1}/{\rm GHz})^{-2}$ K. Substituting $F_\nu = F_{\nu_{\tau=1}}$, $\nu = \nu_{\tau=1}$, $R = 0.18$~pc and the luminosity distance $D = 610$ Mpc for the FRS into the equation above, we obtained a brightness temperature of $\sim5.6\times10^{9}$ K. This temperature is much greater than the physical temperature of the gas (typically $\sim 10^4$ K), which means that the break is not caused by the free-free absorption process.

Another argument against FFA is that the spectral break likely varied with time, from a frequency below 1 GHz to above 1 GHz between 2024 February and late July, and from a high frequency above 1 GHz to below 1 GHz from July to October (Figure \ref{fig:spectral_evolution}). A more natural explanation is that the break frequency is related to 
synchrotron self-absorption (SSA) in an evolving synchrotron nebula.

In the SSA spectral model, the frequency at which the optical depth of the plasma is 1 ($\tau$=1), $\nu_{\tau=1}$, is lower than the observed peak frequency. The redshift of \target~is 0.13, thus we get the spectral break at ${\nu^\prime}_{\rm b} = \nu_{\rm b}(1+z) = 1.76\pm0.23$ GHz in the source rest frame. For a power-law electron distribution with an index $p$ ($1-2\alpha_{\rm high} = 2.46$), we found $\nu_{\tau=1} = 0.8\ {\nu^\prime}_{\rm b} = 1.41\pm0.18$ GHz \citep{1970ranp.book.....P}, while the flux density at $\nu_{\tau=1}$ is derived as $F_{\nu_{\tau=1}} = F_{\nu_{\rm b}} (\nu_{\tau=1}/{\nu^\prime}_{\rm b})^{-0.73\pm0.14} = 171.2\pm29.2\ \mu$Jy. The rest frame SSA frequency and the corresponding flux are used in the following constraints on the plasma of the FRS. 

Similarly, we found $\nu_{\tau=1} = 0.32\ {\nu^\prime}_{\rm b}$ for FRB~20121102A, and applied it to the investigation of the upper limit put forward by the SSA break frequency down to 400 MHz \citep{2021A&A...655A.102R}. The comparison between the regimes allowed by the SSA constraints for the two FRBs, as detailed in the next section, is shown in Figure \ref{fig:B_R_constraints}. 

\section{The size, magnetic field, and minimum energy}

\subsection{The synchrotron self-absorption constraints}
We first derive the dependence of the physical size of the plasma on the magnetic field strength assuming the photons are self-absorbed below the frequency at which $\tau = 1$, at $\nu_{\tau=1}$ the observed flux follows the form of equation~\ref{eq:equation_self_obs}. At a sufficiently low frequency, the brightness temperature $T_{\rm B}$ of any synchrotron source will approach the effective electron temperature $T_e$, see \citet{essential2016} for details. Equating the expressions for $T_e$ (as shown in \citealt{2021A&A...655A.102R}) and Equation~\ref{eq:equation_self_obs}, the observed flux density as a function of the magnetic field strength $B$ \citep{2021A&A...655A.102R} is:
\begin{equation}
F_{\nu}=2.8\times10^{-4}{\rm \mu Jy}\ \left(\frac{\nu_{\tau=1}}{\rm MHz}\right)^{5/2}\ \left(\frac{R_{\rm pc}}{\rm pc}\right)^{2}\ \left(\frac{D_{\rm Gpc}}{\rm Gpc}\right)^{-2}\ \left(\frac{B}{\rm Gauss}\right)^{-1/2}
\label{eq:equation_resmi}
\end{equation}
where $\nu_{\tau=1} = 1.41\pm0.18$ GHz and the modeled flux density is 171.2$\pm$29.2 $\mu$Jy in our case. 

Under the equipartition assumption \citep{essential2016}, for a source with integrated radio luminosity of $L$=$\int_{\nu_{\rm min}}^{\nu_{\rm max}} 4 \pi D^2 F_{\nu} d\nu = 4 \pi D^2 F_{\nu_{\rm max}} \nu_{\rm max}^{-\alpha} [({\nu_{\rm max}^{\alpha+1} - \nu_{\rm min}^{\alpha+1}})/({\alpha+1})]$, which is $\sim 2.4\times10^{38}$\ erg s$^{-1}$ over the entire frequency range, the corresponding magnetic field strength is \citep{essential2016}:
\begin{equation}
B_{\rm min}=[6\sigma_{\rm eq}(1+\eta)c_{12}L]^{2/7}R^{-6/7}
\label{eq:equation_eq_B}
\end{equation}
where $\sigma_{\rm eq} = \epsilon_{B}/\epsilon_{e} = 3/4$ is the ratio of magnetic field energy density to electron energy density and marks the condition of equipartition, $\eta$~is the ratio between the energy of ion and electron and generally we have $\eta \sim$1 since ion synchrotron emission is negligible, $R$ is the source radius, $c_{12}$~is $\sim 1.86\times10^{7}$, calculated with the spectral index and the highest and lowest frequencies observed. Combining Equation~\ref{eq:equation_eq_B} and the $B$ vs.\ $R$ relation obtained from the expression for the self-absorption break, one can derive: 
\begin{equation}
\frac{R_{\rm pc}}{\rm pc}=\left[3.05\times10^{4} \left(\frac{F_{\nu}}{\rm \mu Jy}\right)^{2} \left(\frac{\nu_{\tau=1}}{\rm MHz}\right)^{-5} \left(\frac{D_{\rm Gpc}}{\rm Gpc}\right)^{4}\right]^{7/34}
\label{eq:equation_size_selfabs-and-Bmin}
\end{equation}

We obtain the FRS radius $R_{\rm eq}=0.027\pm0.004~\rm pc$ and the magnetic field strength of $B_{\rm eq}=0.054\pm0.007$ Gauss. The corresponding brightness temperature is $T_{\rm B} \sim4.7\times10^{11}$ K. The minimum total energy of the FRS plasma is written as:
\begin{equation}
E_{\rm min}=c_{13} L^{4/7} R^{9/7}
\label{eq:equation_eq_E}
\end{equation}
where $c_{13} = 0.921\times(c_{12})^{4/7} \sim 1.31\times10^{4}$. The minimum total energy of the FRS is therefore $E_{\rm min} = 6.2\times10^{47}$ ergs. At the current spectral luminosity level, it would take $\sim$ 300 years for the source to emit the total amount of energy; so the existing energy reservoir of the FRS suggests that the radio counterpart will remain for a long time. 

Together with the brightness temperature $T_{\rm B}=4.7\times{10}^{11}$ K of the FRS, we can give the lower limit of the electron density $n_e$ assuming the FRB engine is at the center of the FRS as $n_e\geq\frac{E_{\rm min}}{3kT_e~V}=\frac{E_{\rm min}}{3kT_{\rm B}~V}$, where $V$ is the volume of the FRS as $V=\frac{4}{3}\pi {R}^{3}$. This gives $n_e\geq 1.31~{\rm cm}^{-3}$ and $\rm DM_{src} \geq 0.035 ~pc~{cm}^{-3}$. If the FRB engine is behind the FRS, then the lower limit of DM should be doubled. The reported RM from the bursts is $338.1\pm0.1~{\rm rad}~{\rm m}^{-2}$, which corresponds to a local (redshift-corrected) RM of $\sim (449\pm23)~{\rm rad}~{\rm m}^{-2}$ \citep{2024MNRAS.533.3174T}. Since the RM contributions from the Milky Way galaxy, the intergalactic medium, and probably the FRB host galaxy are small, this value can be regarded as $\rm RM_{src}$. Using the relation $B_\parallel= 1.23\times{10}^{-6}\frac{\rm RM_{src}}{\rm DM_{src}}$ Gauss, we derived the upper limit of the LoS magnetic field strength $B_{\parallel} < 1.6\times{10}^{-2}$ Gauss. This upper limit is smaller than $B_{\rm eq}$ by a factor of 3.4. 

One can derive another independent lower limit of $B_\parallel$ using the burst properties. The measured DM is $527.65\pm0.01~{\rm pc}~{\rm cm}^{-3}$, and the DM contribution from the host galaxy can be estimated as ${333}^{+90}_{-125}~{\rm pc}~{\rm cm}^{-3}$ \citep{2024MNRAS.533.3174T}. If we take this value with redshift correction ($\rm DM_{src}={376.3}~{\rm pc}~{\rm cm}^{-3}$) as the upper limit of $\rm DM_{sr}$, we estimate $B_{\parallel,\rm Burst} > 1.5\times{10}^{-6}$ G. 

We therefore conclude that the  $B_\parallel$ should be in the range of $1.5 \times 10^{-6} {\rm G} < B_\parallel < 1.6 \times 10^{-2} \ {\rm G}$, smaller than than $B_{\rm eq}$. If $B_\parallel$ is close to the upper limit, then a consistent $B$ may be derived if $B$ in the nebula is somewhat below equipartition. For more moderate $B_\parallel$ values, one has to conclude that $B_\parallel \ll B_{\rm eq}$, which may point toward a Poynting-flux-dominated flow. 

\subsection{Physical considerations of the central engine}
The constraints on the magnetic field strength and configuration, as well as the very compact nature of the FRS, reveal the central engine of \target~should be magnetized. The magnetic field configuration with $B_{\rm eq} > B_\parallel$ is consistent with predictions of termination shocks of pulsar wind \citep{2004ApJ...601..479N,2014MNRAS.438..278P} or relativistic jets with a helix magnetic field \citep{1984RvMP...56..255B}. 

The leading FRB engine invokes highly magnetized neutron stars. For a neutron star with a surface magnetic field $B_{\rm NS}$, a spin period of 5 ms, and a radius of 10 km, we can estimate the relation between the magnetic field vs.\ the distance of $R$ away from the neutron star. Inside the light cylinder, the magnetic field strength scales with the radius as $B\propto{R}^{-3}$; outside the light cylinder, the magnetic field strength scales with the radius as $B\propto{R}^{-1}$. In the case of a helix magnetic field of a black hole jet, the same $B\propto{R}^{-1}$ relation applies to all radii. So with the measurements of the magnetic field at a substantial distance from the central engine, such as the magnetic field $B_{\rm eq}$ at the radius $R_{\rm eq}$, one can infer the surface magnetic field strength for a NS engine, and the magnetic field strength at the jey base for a BH engine. 

For a neutron star with a surface magnetic field $B_{\rm S}$, a spin period of $P$, and a radius of $R_{\rm NS}$, we have the light cylinder radius as  
\begin{equation}
R_{\rm LC} = \frac{cP}{2\pi}\approx2.4\times{10}^{2}~{\rm km}~(\frac{P}{5~\rm ms})
\label{eq:equation_RLC}
\end{equation}
and the magnetic field strength at the light cylinder
\begin{equation}
B_{\rm LC} =B_{\rm S} {\left(\frac{R_{\rm LC}}{R_{\rm NS}}\right)}^{-3}\approx7.4\times{10}^{7}~{\rm G}~\left(\frac{B_{\rm S}}{{10}^{12}~\rm G}\right){\left(\frac{P}{5~\rm ms}\right)}^{-3}{\left(\frac{R_{\rm NS}}{10~\rm km}\right)}^{3}
\label{eq:equation_BLC}
\end{equation}
Outside the light cylinder, we have the $B_{\rm R}$ vs.\ $R$ relation as
\begin{equation}
B_{\rm R} =B_{\rm LC}{\left(\frac{R}{R_{\rm LC}}\right)}^{-1}\approx5.7\times{10}^{-2}~{\rm G}~{\left(\frac{B_{\rm S}}{{10}^{12}~\rm G}\right)} {\left(\frac{P}{5~\rm ms}\right)}^{-2} {\left(\frac{R_{\rm NS}}{10~\rm km}\right)}^{3} {\left(\frac{R}{0.01~\rm pc}\right)}^{-1}
\label{eq:equation_BR1}
\end{equation}
We found the FRS with a magnetic field strength $B_{\rm eq}=0.054~\rm G$ at the radius $R=0.027~\rm pc$ corresponds to a magnetic neutron star engine with the surface magnetic field strength of $B_{\rm NS}= 2.56\times{10}^{12}~\rm G$ for the neutron star with a spin period of 5 ms and a radius of 10 km. 

Using these numbers, we can obtain a general relation between the magnetic field strength and the radius that follows the relation $B\propto R^{-1}$, corresponding to the derived $B_{\rm eq}$ and $R_{\rm eq}$
\begin{equation}
B_{\rm R} \approx0.15~{\rm G}~{\left(\frac{R}{0.01~\rm pc}\right)}^{-1}
\label{eq:equation_BR2}
\end{equation}

Assume the upstream magnetic field strength $B\propto R^{-1}$ for a black hole mass of $M_{\rm BH}$, we obtain the magnetic field at the jet base at $6~r_{g}$ as

\begin{equation}
B_{\rm R} \approx5.08\times{10}^{9}~{\rm G}~{\left(\frac{R}{6~r_{\rm g}}\right)}^{-1}~{\left(\frac{M_{\rm BH}}{1~M_{\odot}}\right)}^{-1}.
\label{eq:equation_BR4}
\end{equation}
A steeper dependence of the magnetic field strength at the jet base on radius would require an even larger magnetic field at smaller radii. 

Therefore, to produce the magnetic field strength $B_{\rm eq}$ at the radius $R_{\rm eq}$ from the central engine, as determined and constrained in the FRS, a highly magnetized central engine is required for both neutron star and black hole scenarios. 

%\subsection{Constraints on the Magnetic Field Strength Constraints}
\section{The flaring radio source as a slow radio transient}

Different types of slow radio transients, such as those associated with GRBs, SNe, TDEs, magnetar flares, AGN/Blazars, and XRBs, occupy different regimes in the parameter space of the flare peak luminosity, the e-folding rise or decay time scale,  and the corresponding frequency of the flux peak \citep{2015MNRAS.446.3687P,2024NatAs...8.1159C}. The location of the radio flare observed in the FRS can potentially reveal the origin of the FRB by associations with known source types. 

The sensitive radio observations likely missed the actual luminosity peak of the radio flare. The flare peak probably occurred between the MeerKAT observation in 2024 February and the VLA observations in 2024 late July, as indicated by an apparent steeper decay from the linear fit to the light curve (Figure \ref{fig:lc_spec}). We were able to measure the e-folding rise and decline time scales as $\tau_{r}=235\pm76$ days and $\tau_{d}=168\pm95$ days, respectively, with the corresponding observed peak flux density as 131$\pm$10 $\mu$Jy at 1.5 GHz. As shown in Figure \ref{fig:comparison_transients_w} and Figure \ref{fig:comparison_transients_efolding}, the FRS is located nearest to the radio SNe sample and the LLAGN sample. Specifically, the FRS has a spectral luminosity comparable to those of the brightest radio SNe but with an e-folding time scale longer than those SNe Ia/b but shorter than the SNe IIn such as SN 1988z. The nearest LLAGN sample corresponds to the measurements of radio flares in NGC 7213. 

With the detection of the SSA in the FRS, we can compare the FRS with those SNe as well as the model for radio SNe (with the spectral index of 2.5) \citep{1998ApJ...499..810C,2012ApJ...752...78S}. This is shown in Figure \ref{fig:FRS_sne}. If the FRB corresponds to a newborn source, its radio flare behavior is mostly close to the engine-powered SNe; if the FRB corresponds to an FRB of half years to three years old when it was discovered, its radio flare should have a luminosity substantially higher than those SN Ia/b; our study does not support potential link of the radio flare of the FRS to those of Super-Luminous SNe (SLSNe). The similarity to engine-powered SNe and accretion-powered LLAGN strongly suggests that the central engine activities should have played a significant role in generating the FRS flare. 

\section{The link to the PRSs associated with FRBs}

In addition to the radio flare which shows the distinction from those PRSs, all the radio spectra of the FRS in the high-frequency range ($>$1 GHz) were consistent with a steep spectrum, with the power-law indices measured as -1.1$\pm$0.8, -0.73$\pm$0.14, and -0.43$\pm$0.21, respectively, in 2024 February, July, and October. On the other hand, non-detections as well as the detection at the low-frequency end($<$1 GHz) \citep{2024ATel16452....1K,2024arXiv240509749P,2024arXiv240612804K,2024ATel16820....1B,2024arXiv241213121B} implies a turn-over at the lower frequencies (Figure \ref{fig:spectral_evolution}), indicating an evolving break frequency due to SSA during the flare. Such an SSA spectral break has never been detected in any of those PRSs or candidates before. Based on the relation between the radio flux density, source size, and magnetic field in the SSA calculation shown above, the rise in the radio flux density between 2024 February and July likely corresponded to an increase in both the magnetic field strength and the size with an increasing SSA frequency, while the decline in the radio flux density between 2024 July and October likely corresponded to a decrease of the magnetic field or an increase of the source size with a decreasing SSA frequency. The optically thin part of the radio spectra are all steeper than those of compact PRSs \citep{2023ApJ...959...89Z}. We are not able to determine if a cooling break existed due to the sensitivity limit. 

\begin{figure}[H]
 \centering
 \includegraphics[height=7.0in]{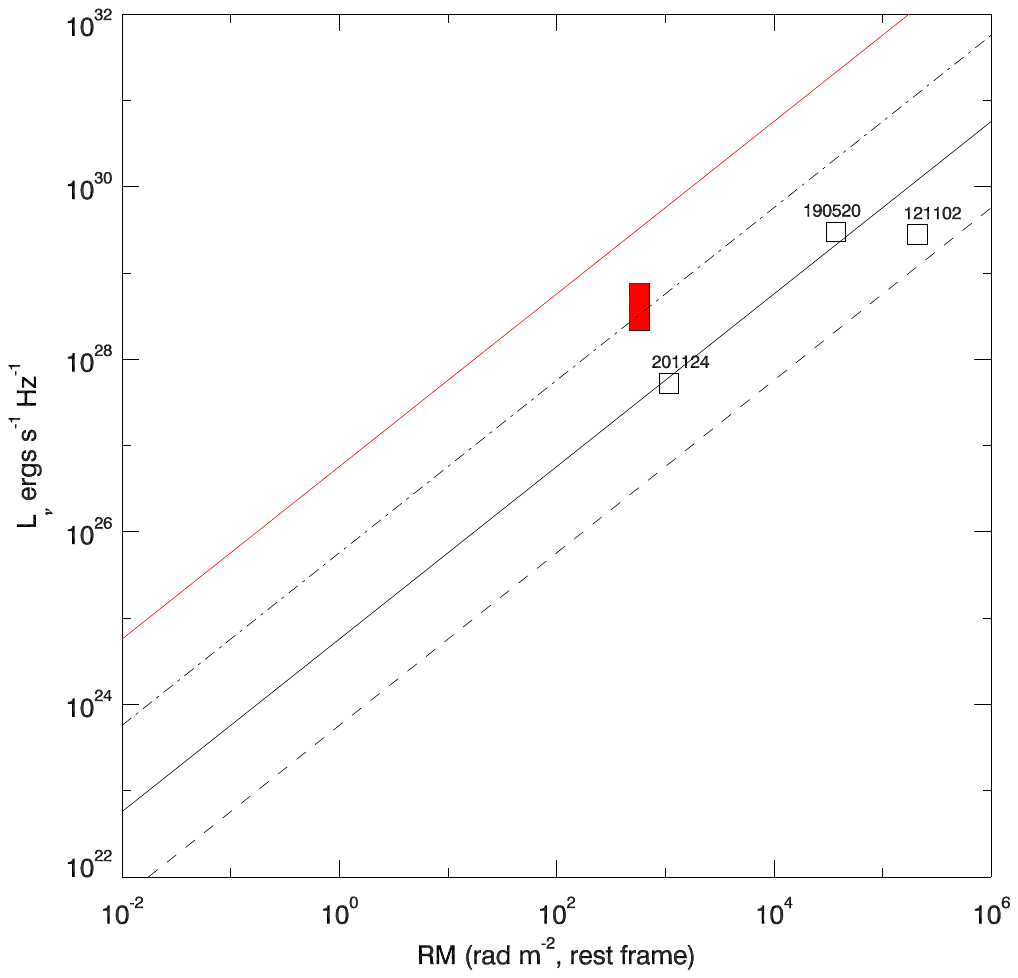}
 \vspace{-2.0cm}
 \caption{
 \footnotesize \textbf{The spectral luminosity and the rotation measure for the FRS and the PRSs.} Spectral luminosity measurements with observations in 2024 February, July, and October for the FRS are shown as three overlapping red squares. The measurements of the PRSs are shown as open black squares. The actual peak of the radio flare was likely missed during the period between early February and late July, thus the FRS could have reached a larger spectral luminosity. The FRS had an observed peak luminosity going beyond the predicted spectral luminosity range corresponding to the parameter ${\rm p}_{\rm n}=\zeta_{e}\gamma {c}^{2}(R/0.01~\rm pc)^{2}$ in the range [0.1,10] (dashed lines), which is defined by those PRSs and candidates \citep{2024Natur.632.1014B,2024ApJ...976..199I}. The black and red solid lines correspond to the relations of ${\rm p}_{\rm n}$ = 1 and 100, respectively.}
 \label{fig:spec_L-RM}
\end{figure}

\begin{figure}[H]
 \centering
 \includegraphics[height=7.0in]{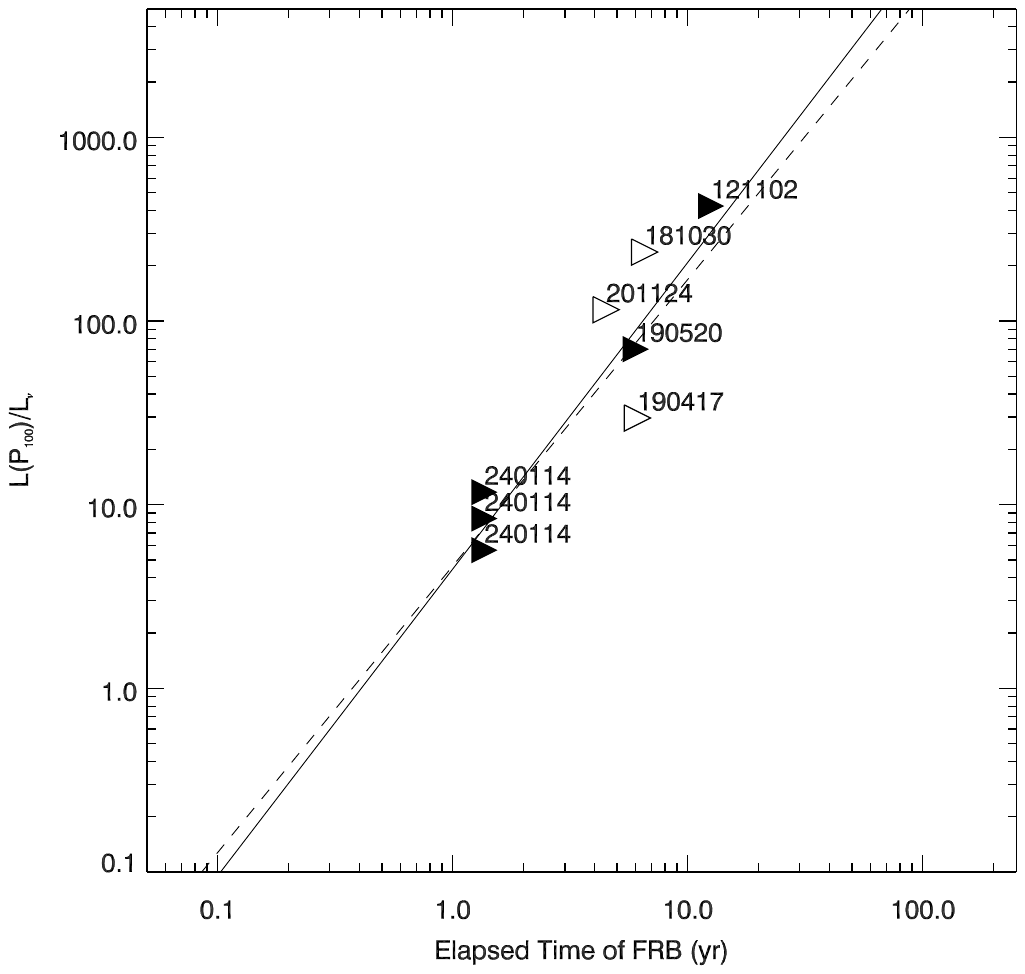}
 \vspace{-2.0cm}
 \caption{
 \footnotesize \textbf{A likely age indicator of the FRBs with a radio continuum counterpart.} The spectral luminosity ratio between that corresponding to the nebular model of ${\rm p}_{\rm n} =100$ (red line in Figure \ref{fig:spec_L-RM}) and the observed value is correlated with the elapsed time of an FRB since its discovery for the FRS and the PRSs and candidates. The FRS and two most compact ($<$ 10 pc) PRSs are marked as solid right-facing arrows. An additional shift of 0.5 years to later time is applied to the elapsed time for all the sources, to potentially account for that \target~was half a year old when it was discovered. The nebular model parameter ${\rm p}_{\rm n}=\zeta_{e}\gamma {c}^{2}(R/0.01~\rm pc)^{2}$ \citep{2024Natur.632.1014B}. The best-fit relations corresponding to (1) the FRS and the two compact PRSs (solid line) and (2) all sources (dashed line) are shown. The positive correlation indicates the spectral luminosity ratio (vertical axis) for an FRB can serve as an empirical indicator of the FRB age, at least for younger FRBs, since it probably indicates central engine activities.
 }
 \label{fig:elapsed_time}
\end{figure}

The similarity of the radio flare to those seen in SNe and LLAGNs and the likely evolving SSA spectra indicate central engine activities. \target~has also shown very long active episode of very high burst rate \citep{2024ATel16505....1Z} and coincident gamma-ray flares \citep{2024arXiv241106996X}, which has never been seen in other FRBs. All these properties imply that \target~is much younger than those FRBs found associated with a PRS previously. On the other hand, it has been predicted that there should be a positive correlation between the PRS luminosity and the RM in the nebular model of the PRS \citep{2020ApJ...895....7Y,2022ApJ...928L..16Y,2024Natur.632.1014B}. But even for a nebula of a radius of a few pc, it would need years or longer to form, as the expansion velocity is at most a fraction of the light speed. Therefore, any deviation from the nebular model prediction would be an indicator of the central engine activity instead, which might be used to diagnose newborn FRBs. In the FRS case, it indeed showed a luminosity excess beyond the nebular model as compared with other sources (Figure \ref{fig:spec_L-RM}). We found that there is a correlation between the deviation in the spectral luminosity from the nebular model prediction of the spectral luminosity with the parameter ${\rm p}_{\rm n}=\zeta_{e}\gamma {c}^{2}(R/0.01~\rm pc)^{2}$ being 100 and the elapsed time defined as the time interval since the discovery of the corresponding FRB, for the FRS, the PRSs and candidates (Figure \ref{fig:elapsed_time}). The elapsed time of each FRB is the proximity of the lower limit of the age of an FRB. It could be a fraction of the actual age for sources like FRB~20121102A since there had been no sensitive monitoring before CHIME operated. In the case of \target, since it was discovered by a wide field-of-view instrument CHIME with only 4-minutes daily time window, the actual birth time of the FRB engine is very likely at least weeks or months before the first CHIME detection, as the time separation between the first and the second burst is about a week. To account for the inaccuracy in the elapsed time as the proxy of the age for \target~(see also Figure \ref{fig:lc_spec}), we added a universal shift of 0.5 years to all sources in Figure \ref{fig:elapsed_time}. We found a change in the universal shift within a few years will not change the result significantly. The correlation we found indicates that the spectral luminosity ratio between that in the nebular model of ${\rm p}_{\rm n}=100$ and the observed spectral luminosity can be an empirical indicator of the age of the corresponding FRB, probably because the younger the FRB, the more injection of energy by the central engine, which contribute to extra spectral luminosity due to extra shock or outflow other than that of the nebulae in the FRSs. 

The total energy of the FRS is larger than the minimum energy $E_{\rm min} = 6.2\times10^{47}$ ergs constrained from equipartition. It suggests that the radio source can remain at the current flux density level for hundreds of years, although the SSA and potential synchrotron cooling would modulate its exact manifestations as a radio source. As the central engine gets less energetic, we speculate that the energy injection and activity will be reduced, and then the source will likely behave and show as a PRS.  

\section{Constraints from multi-wavelength observations}
\begin{figure}[H]
 \centering
 \includegraphics[angle=0,height=7.5in]{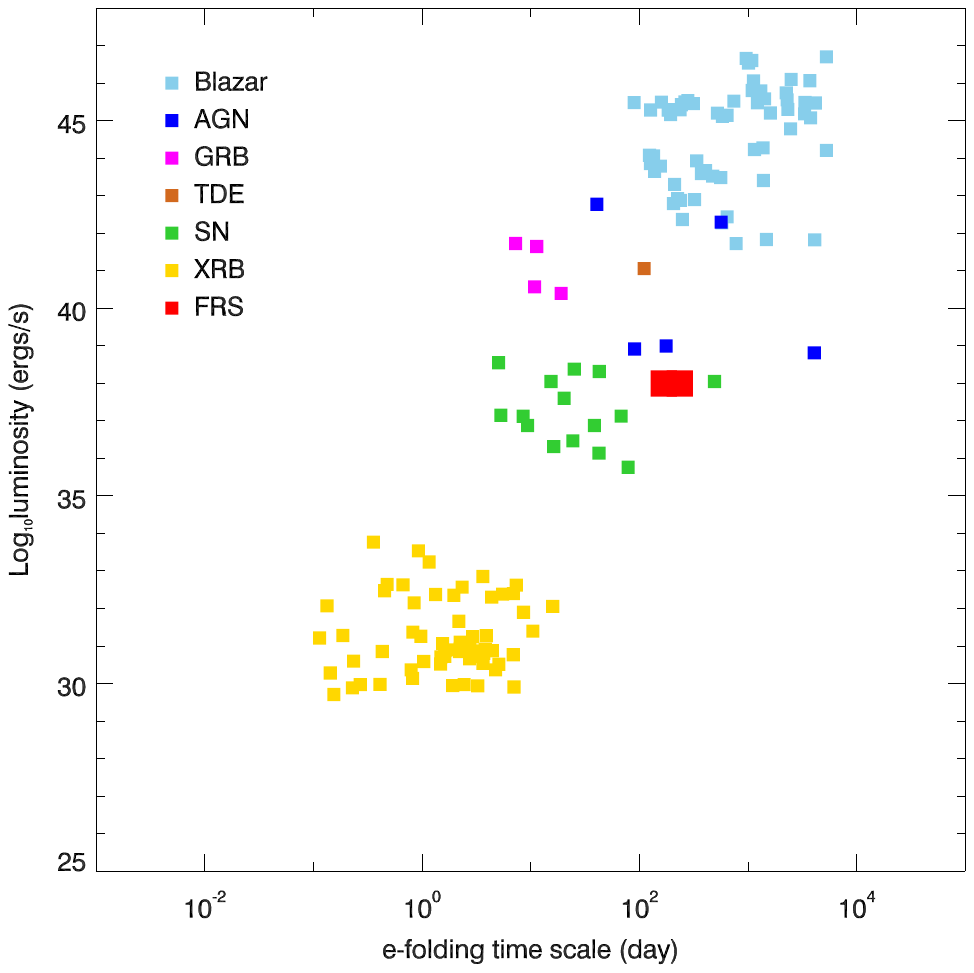}
 \vspace{-2.0cm}
 \caption{
 \footnotesize \textbf{The relation between the radio luminosity and the rest-frame e-folding time scale for several different types of radio transients.} The data for SNe, GRBs, XRBs, TDE, AGN/Blazars are from previous samples \citep{2015MNRAS.446.3687P,2024NatAs...8.1159C}. The two overlapping red squares represent the measurements of the e-folding times of the radio flare of the FRS. The SN sample to the right of \target~is SN 1988Z, an SN IIn. The AGN samples nearest to the FRS are from the LLAGN NGC 7213.}
 \label{fig:comparison_transients_efolding}
\end{figure}

In addition to the detection of the gamma-ray flares coincident with the activation of \target~\citep{2024arXiv241106996X}, which is suggestive of central engine activities of a newborn source, we also searched for potential multi-wavelength counterparts, as the FRS independently showed similarity to SNe and LLAGNs in the peak luminosity vs.\ variability time scale plane (Figures \ref{fig:comparison_transients_w}, \ref{fig:FRS_sne} and \ref{fig:comparison_transients_efolding}). No X-ray nor optical counterpart of \target~have been found in previous efforts with $Swift$ \citep{2024ATel16645....1V}; our VLA observations also triggered VLA/Realfast and subsequent $Swift$ follow-ups with high time-resolution observations. We made use of the $Swift$ data in the period between MJD 60437 and 60522 to put a further constraint on the multi-wavelength, persistent counterpart of the FRS, some of which were performed simultaneously with the VLA observations in 2024 July. We did not achieve any significant detections of X-ray, UV and optical counterparts. We then stacked the six $Swift$/XRT observations available in the Photon Counting (PC) mode and obtained a 3 $\sigma$ upper limit of $1.43 \times 10^{-3} $c/s. By assuming a power-law X-ray spectrum with photon index 2, the 3 $\sigma$ flux upper limit in the 0.5--10 keV band is $4.78\times 10^{-14} $ erg s$^{-1}$ cm $^{-2}$ for a Galactic HI column density $n_{H}=4.98 \times 10^{20}$ cm$^2$, which corresponds to an X-ray luminosity $2.13\times10^{42}$ erg s$^{-1}$ at a distance of $\sim$ 610 Mpc. There are five $Swift$/UVOT observations performed in the U filter and 12 observations in the UVM2 filter. The total exposure times are 2528 and 9299 seconds, respectively. We stacked the images in each filter and obtained the 3 $\sigma$ upper limit 22.60 and 23.89 in AB mag, which corresponds to the luminosity as $1.28\times10^{42}$ erg s$^{-1}$ in the U band and $6.11\times10^{41}$ erg s$^{-1}$ in the UVM2 band. The $Swift$ observations set the upper limit on the X-ray and optical/UV luminosity from a potential LLAGN or an intermediate-mass or stellar-mass black hole when the FRS  occurred and \target~has been active since early 2024. 

\begin{figure*}
\centering
\includegraphics[width=0.85\textwidth,angle=0]{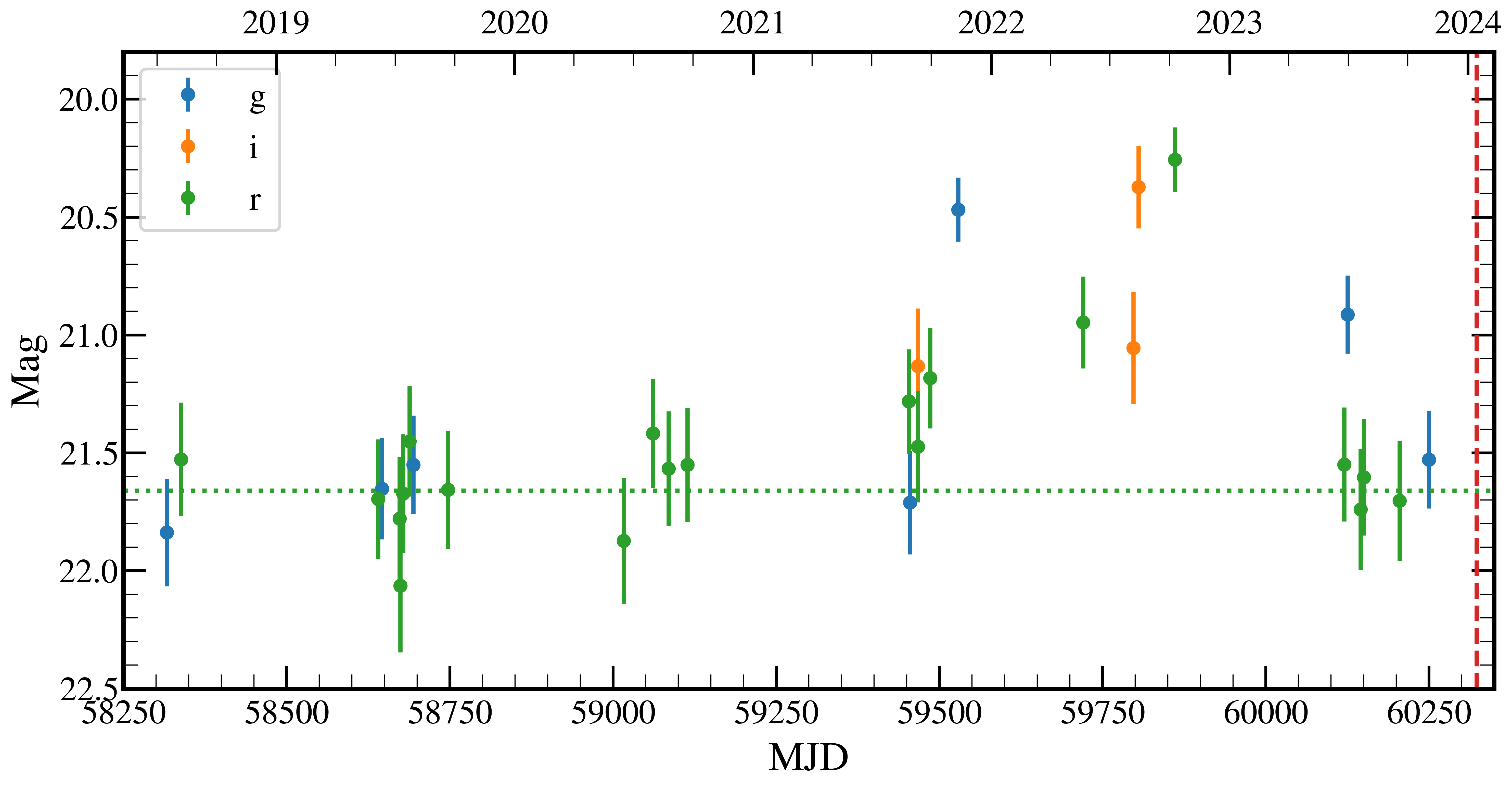}

\caption{
\footnotesize \textbf{The ZTF light curve of SDSS J212739.84$+$041945.8.} The host dwarf galaxy had an optical/infrared flare during the period between 2021 and 2023. Since the second half of 2023, the dwarf galaxy returned to the normal, quiescent magnitude (green dotted line), about half a year before \target~was discovered by CHIME (the vertical red dashed line).}
\label{fig:ztf}
\end{figure*}

The host galaxy of \target~is identified as SDSS J212739.84$+$041945.8 \citep{2024MNRAS.533.3174T,2024ATel16446....1T,2024ATel16613....1B}. We found evidence of activity in the host in the recent past, about half a year before \target~was active. We searched the ZTF archive by using a Python package \textsc{ztfquery} \citep{mickael_rigault_2018} and obtained the light curve in the period of 2019--2023 (\autoref{fig:ztf}). The observation coverage at the $r$ band is the best. It appears that the galaxy remains quiescent before 2021 at a brightness level $\sim$ 21.66 mag at the $r$ band. There was an outburst or flare during the period 2022-2023, during which the brightness at the $r$ band roughly increased by 1.4 Mag in a period of $\sim$ 400 days. The peak luminosity of the outburst at the $r$ band is roughly 7$\times10^{42}$ erg s$^{-1}$, which is lower than that of SLSNe ($>10^{43}$ erg s$^{-1}$) \citep{Gal-Yam2012Sci}. The total energy from integrating the outburst is roughly 3$\times 10^{50}$ ergs, which is also smaller than that of a typical SLSN ($10^{51}$ ergs) but significantly larger than those of X-ray binary outbursts. The rising timescale of the flare was not well-constrained, probably longer than that seen in a typical SLSN ($\lesssim$ 50 days). We can not rule out the possibility that a potential normal SN had occurred in the host in the period about half a year to three years before the activation of \target~in early 2024, but the possibility of a SLSN can be ruled out based on the luminosity limit. If the FRB engine was half years to three years old (due to a past SN) when \target~was discovered, since the FRS is significantly brighter in radio than the SNe sample as seen in Figure \ref{fig:FRS_sne}, the central engine should have been active. Sensitive optical/Infrared observations can put more stringent constraints. 

%\section{Supplementary text}
%\subsection{Contour Images of the Detected 24 Discrete Blobs}
%\subsection{Classifications of Discrete Blobs and Alternative Model Fits}

%\section{Supplementary text}
%\subsection{Contour Images of the Detected 24 Discrete Blobs}
%\subsection{Classifications of Discrete Blobs and Alternative Model Fits}

\clearpage

%\section{Gold Open Access}

%\section{Author publication charges} \label{sec:pubcharge}

%\section{Rotating tables} \label{sec:rotate}

%\section{Using Chinese, Japanese, and Korean characters}


%%%%%%%Supplementary figures%%%%%%%%%%

%\begin{figure*}
%\centering
%\includegraphics[width=0.80\textwidth,angle=0]{VLA202407_MeerKAT.pdf}

%\caption{
%\footnotesize \textbf{Flux densities of the FRS and cross matched point sources.} Flux densities of the cross matched 17 ``point" sources and the FRS measured with VLA observations taken in late
%July, and MeerKAT observations taken in February, respectively. }
%\label{fig:vla07_meerkat}
%\end{figure*}
%\clearpage

\clearpage

\clearpage

%\begin{figure*}%[H]
%  \centering
%  \subfloat{
%        \includegraphics[width=0.55\textwidth]{VLA202407_MeerKAT.pdf}
%    }
    %\hspace{0.1cm}
%    \subfloat{
%        \includegraphics[width=0.55
%        \textwidth]{VLA202407_10.pdf}
%    }
%  \caption{\footnotesize \textbf{Flux densities of the FRS and cross matched point sources.} \textbf{A}. Flux densities of the cross matched 17 ``point" sources and the FRS measured with VLA observations taken in late July, and MeerKAT observations taken in February, respectively. \textbf{B}. Flux densities of the cross matched 15 ``point" sources and the FRS measured with VLA observations taken in October and July, respectively.}
%  \label{fig:twoplots}
%\end{figure*}

\clearpage

\clearpage

\clearpage

%\begin{figure*}
%\centering
%\includegraphics[width=0.85\textwidth,angle=0]{ZTF.png}
%
%\caption{
%\footnotesize \textbf{The ZTF light curve of SDSS J212739.84+041945.8.}
%Since the second half of 2023, the dwarf galaxy returned normal magnitudes.}
%\label{fig:ztf}
%\end{figure*}
%\clearpage

%%%%%%%%%Supplementary tables%%%%%%%%% 
\newpage
%\begin{sidewaystable}
\begin{table}
\rmfamily
\centering
\footnotesize
\begin{tabular}{|l|c|c|c|c|c|c|}
\hline
\bf Date & \bf MJD & \bf Telescope & \bf Frequency & \bf Beam size & \bf Beam position angle & \bf Flux density\\
&&& \bf (GHz) & \bf ($^{\prime\prime}$,$^{\prime\prime}$) & (\bf $\arcdeg$) & \bf ($\mu$Jy) \\
\hline 
2024 Feb 09 & 60349 & MeerKAT & 0.8 &
26.15$\times$12.31 & 50.98 & $<$120 \\
2024 Feb 09 & 60349 & MeerKAT & 1.5 &
8.63$\times$6.24 & 11.34 & 64$\pm$14 \\
2024 July 23 & 60514 & VLA & 1.5 &
3.51$\times$3.18 & -28.06 & 123$\pm$15 \\
2024 July 27 & 60518 & VLA & 1.5 &
4.24$\times$3.21 & 54.00 & 101$\pm$17 \\
2024 July 27 & 60518 & VLA & 3 &
1.96$\times$1.73 & 76.79 & 91$\pm$10 \\
2024 July 27 & 60518 & VLA & 5.5 & 
1.30$\times$1.03 & 88.84 & 53$\pm$9 \\
2024 July 29 & 60520 & VLA & 1.5 &
4.22$\times$3.26 & 58.18 & 124$\pm$16 \\
2024 July 29 & 60520 & VLA & 3 &
2.28$\times$1.80 & 62.56 & 70$\pm$11 \\
2024 July 29 & 60520 & VLA & 5.5 & 
1.43$\times$1.04 & 85.78 & 74$\pm$10 \\
2024 Oct 03 & 60586 & VLA & 1.5 & 
5.02$\times$1.22 & -72.83 & 87$\pm$19 \\
2024 Oct 03 & 60586 & VLA & 3 & 
2.78$\times$0.58 & -69.21 & 72$\pm$12 \\
2024 Oct 03 & 60586 & VLA & 5.5 & 
1.87$\times$0.34 & -66.48 & 50$\pm$9 \\

\hline 
\end{tabular} 

\caption{\footnotesize \textbf{Summary of the MeerKAT and the VLA observations and measurements in 2024.} The frequency refers to the central frequency of the observing band, except that the central frequency in the MeerKAT L-band refers to the high sub-band central frequency.} 
\label{tab:meerkat+vla_observations}
\end{table}

\begin{table}
\rmfamily
\centering
\footnotesize
\begin{tabular}{l c c c c c}
\hline

Phase & \bf $m_{\rm p}$ & \bf $t_{\rm r}$ & \bf $\theta_{\rm r}$ & $m_{\rm o}'$ & $\chi^2$ (dof) \\
& \bf{} & (Day) & $({\mu as})$ & \bf{} & \\
%scattering disk 10 kpc away

%20240214, 202407vla
\hline
Rise & $\sim~32.7\%$ & $\sim 6.4$ & $\sim190$ & 48.6$\pm$14.4\% & 15.2 (1)  \\
Decline & $\sim~32.7\%$ & $\sim 6.4$ & $\sim190$ & 28.5$\pm$15.7\% & 4.2 (1)  \\

%\hline(20240214,20240723,27,29,20241003)
%1.5 & $\sim~32.7\%$ & $\sim 6.4$ & $\sim190$ & 25.4$\pm$7.8\% & 2.9 (4)  \\

%\hline(20240214,202407,20241003)
%1.5 & $\sim~32.7\%$ & $\sim 6.4$ & $\sim190$ & 34.9$\pm$12.0\% & 4.6 (2)  \\ 

\hline 
\end{tabular} 

\caption{\footnotesize \textbf{Summary of scintillation analysis for the FRS.} The observed flux variation coefficients and variability significances of the FRS at 1.5~GHz corresponding to the rise phase and decline phase, respectively, are shown. Also shown here are the predicted Galactic refractive scintillation properties of the FRS, assuming a scattering disk that is 10~kpc away from us, including the size of the scattering disk and variability timescale.}
\label{ta:predictions}
\end{table} 

\clearpage

\bibliography{FRB20240114A}{}

\begin{thebibliography}{}
\expandafter\ifx\csname natexlab\endcsname\relax\def\natexlab#1{#1}\fi
\providecommand{\url}[1]{\href{#1}{#1}}
\providecommand{\dodoi}[1]{doi:~\href{http://doi.org/#1}{\nolinkurl{#1}}}
\providecommand{\doeprint}[1]{\href{http://ascl.net/#1}{\nolinkurl{http://ascl.net/#1}}}
\providecommand{\doarXiv}[1]{\href{https://arxiv.org/abs/#1}{\nolinkurl{https://arxiv.org/abs/#1}}}

\bibitem[{{Abbott} {et~al.}(2023){Abbott}, {Abbott}, {Acernese}, {Ackley},
  {Adams}, {Adhikari}, {Adhikari}, {Adya}, {Affeldt}, {Agarwal}, {Agathos},
  {Agatsuma}, {Aggarwal}, {Aguiar}, {Aiello}, {Ain}, {Ajith}, {Akutsu},
  {Albanesi}, {Allocca}, {Altin}, {Amato}, {Anand}, {Anand}, {Ananyeva},
  {Anderson}, {Anderson}, {Ando}, {Andrade}, {Andres}, {Andri{\'c}},
  {Angelova}, {Ansoldi}, {Antelis}, {Antier}, {Appert}, {Arai}, {Arai}, {Arai},
  {Araki}, {Araya}, {Araya}, {Areeda}, {Ar{\`e}ne}, {Aritomi}, {Arnaud},
  {Aronson}, {Arun}, {Asada}, {Asali}, {Ashton}, {Aso}, {Assiduo}, {Aston},
  {Astone}, {Aubin}, {Austin}, {Babak}, {Badaracco}, {Bader}, {Badger}, {Bae},
  {Bae}, {Baer}, {Bagnasco}, {Bai}, {Baiotti}, {Baird}, {Bajpai}, {Ball},
  {Ballardin}, {Ballmer}, {Balsamo}, {Baltus}, {Banagiri}, {Bankar},
  {Barayoga}, {Barbieri}, {Barish}, {Barker}, {Barneo}, {Barone}, {Barr},
  {Barsotti}, {Barsuglia}, {Barta}, {Bartlett}, {Barton}, {Bartos}, {Bassiri},
  {Basti}, {Bawaj}, {Bayley}, {Baylor}, {Bazzan}, {B{\'e}csy}, {Bedakihale},
  {Bejger}, {Belahcene}, {Benedetto}, {Beniwal}, {Bennett}, {Bentley},
  {Benyaala}, {Bergamin}, {Berger}, {Bernuzzi}, {Berry}, {Bersanetti},
  {Bertolini}, {Betzwieser}, {Beveridge}, {Bhandare}, {Bhardwaj},
  {Bhattacharjee}, {Bhaumik}, {Bilenko}, {Billingsley}, {Bini}, {Birney},
  {Birnholtz}, {Biscans}, {Bischi}, {Biscoveanu}, {Bisht}, {Biswas}, {Bitossi},
  {Bizouard}, {Blackburn}, {Blair}, {Blair}, {Blair}, {Bobba}, {Bode}, {Boer},
  {Bogaert}, {Boldrini}, {Bonavena}, {Bondu}, {Bonilla}, {Bonnand}, {Booker},
  {Boom}, {Bork}, {Boschi}, {Bose}, {Bose}, {Bossilkov}, {Boudart},
  {Bouffanais}, {Boumerdassi}, {Bozzi}, {Bradaschia}, {Brady}, {Bramley},
  {Branch}, {Branchesi}, {Brau}, {Breschi}, {Briant}, {Briggs}, {Brillet},
  {Brinkmann}, {Brockill}, {Brooks}, {Brooks}, {Brown}, {Brunett}, {Bruno},
  {Bruntz}, {Bryant}, {Buchanan}, {Bulik}, {Bulten}, {Buonanno}, {Buscicchio},
  {Buskulic}, {Buy}, {Byer}, {Cadonati}, {Cagnoli}, {Cahillane}, {Bustillo},
  {Callaghan}, {Callister}, {Calloni}, {Cameron}, {Camp}, {Canepa},
  {Canevarolo}, {Cannavacciuolo}, {Cannon}, {Cao}, {Cao}, {Capocasa}, {Capote},
  {Carapella}, {Carbognani}, {Carlin}, \& {Carney}}]{LVK23}
{Abbott}, R., {Abbott}, T.~D., {Acernese}, F., {et~al.} 2023, \apj, 955, 155,
  \dodoi{10.3847/1538-4357/acd770}

\bibitem[{{Andersson} {et~al.}(2023){Andersson}, {Lintott}, {Fender}, {Bright},
  {Carotenuto}, {Driessen}, {Espinasse}, {Gasealahwe}, {Heywood}, {van der
  Horst}, {Motta}, {Rhodes}, {Tremou}, {Williams}, {Woudt}, {Zhang}, {Bloemen},
  {Groot}, {Vreeswijk}, {Giarratana}, {Saikia}, {Andersson}, {Ruiz Arroyo},
  {Baert}, {Baumann}, {Domainko}, {Eschweiler}, {Forsythe}, {Gaudenzi}, {Ann
  Grenier}, {Iannone}, {Lahoz}, {Melville}, {De Sousa Nascimento}, {Navarro},
  {Parthasarathi}, {Piilonen}, {Rahman}, {Smith}, {Stewart}, {Temoke},
  {Tworek}, \& {Whittle}}]{2023MNRAS_meerkat}
{Andersson}, A., {Lintott}, C., {Fender}, R., {et~al.} 2023, \mnras, 523, 2219,
  \dodoi{10.1093/mnras/stad1298}

\bibitem[{{Anna-Thomas} {et~al.}(2023){Anna-Thomas}, {Connor}, {Dai}, {Feng},
  {Burke-Spolaor}, {Beniamini}, {Yang}, {Zhang}, {Aggarwal}, {Law}, {Li},
  {Niu}, {Chatterjee}, {Cruces}, {Duan}, {Filipovic}, {Hobbs}, {Lynch}, {Miao},
  {Niu}, {Ocker}, {Tsai}, {Wang}, {Xue}, {Yao}, {Yu}, {Zhang}, {Zhang}, {Zhu},
  \& {Zhu}}]{2023Sci...380..599A}
{Anna-Thomas}, R., {Connor}, L., {Dai}, S., {et~al.} 2023, Science, 380, 599,
  \dodoi{10.1126/science.abo6526}

\bibitem[{{Begelman} {et~al.}(1984){Begelman}, {Blandford}, \&
  {Rees}}]{1984RvMP...56..255B}
{Begelman}, M.~C., {Blandford}, R.~D., \& {Rees}, M.~J. 1984, Reviews of Modern
  Physics, 56, 255, \dodoi{10.1103/RevModPhys.56.255}

\bibitem[{{Bhandari} {et~al.}(2023){Bhandari}, {Marcote}, {Sridhar},
  {Eftekhari}, {Hessels}, {Hewitt}, {Kirsten}, {Ould-Boukattine}, {Paragi}, \&
  {Snelders}}]{2023ApJ...958L..19B}
{Bhandari}, S., {Marcote}, B., {Sridhar}, N., {et~al.} 2023, \apjl, 958, L19,
  \dodoi{10.3847/2041-8213/ad083f}

\bibitem[{{Bhardwaj} {et~al.}(2024){Bhardwaj}, {Kirichenko}, \& {Gil de
  Paz}}]{2024ATel16613....1B}
{Bhardwaj}, M., {Kirichenko}, A., \& {Gil de Paz}, A. 2024, The Astronomer's
  Telegram, 16613, 1

\bibitem[{{Bhattacharya} {et~al.}(2024){Bhattacharya}, {Murase}, \&
  {Kashiyama}}]{2024arXiv241219358B}
{Bhattacharya}, M., {Murase}, K., \& {Kashiyama}, K. 2024, arXiv e-prints,
  arXiv:2412.19358, \dodoi{10.48550/arXiv.2412.19358}

\bibitem[{{Bhusare} {et~al.}(2024{\natexlab{a}}){Bhusare}, {Maan}, \&
  {Kumar}}]{2024ATel16820....1B}
{Bhusare}, Y., {Maan}, Y., \& {Kumar}, A. 2024{\natexlab{a}}, The Astronomer's
  Telegram, 16820, 1

\bibitem[{{Bhusare} {et~al.}(2024{\natexlab{b}}){Bhusare}, {Maan}, \&
  {Kumar}}]{2024arXiv241213121B}
---. 2024{\natexlab{b}}, arXiv e-prints, arXiv:2412.13121,
  \dodoi{10.48550/arXiv.2412.13121}

\bibitem[{{Bochenek} {et~al.}(2020){Bochenek}, {Ravi}, {Belov}, {Hallinan},
  {Kocz}, {Kulkarni}, \& {McKenna}}]{STARE2-SGR}
{Bochenek}, C.~D., {Ravi}, V., {Belov}, K.~V., {et~al.} 2020, \nat, 587, 59,
  \dodoi{10.1038/s41586-020-2872-x}

\bibitem[{{Bruni} {et~al.}(2024{\natexlab{a}}){Bruni}, {Piro}, {Yang}, {Quai},
  {Zhang}, {Palazzi}, {Nicastro}, {Feruglio}, {Tripodi}, {O'Connor}, {Gardini},
  {Savaglio}, {Rossi}, {Nicuesa Guelbenzu}, \&
  {Paladino}}]{2024Natur.632.1014B}
{Bruni}, G., {Piro}, L., {Yang}, Y.-P., {et~al.} 2024{\natexlab{a}}, \nat, 632,
  1014, \dodoi{10.1038/s41586-024-07782-6}

\bibitem[{{Bruni} {et~al.}(2024{\natexlab{b}}){Bruni}, {Piro}, {Yang},
  {Palazzi}, {Nicastro}, {Rossi}, {Savaglio}, {Maiorano}, \&
  {Zhang}}]{2024arXiv241201478B}
{Bruni}, G., {Piro}, L., {Yang}, Y.~P., {et~al.} 2024{\natexlab{b}}, arXiv
  e-prints, arXiv:2412.01478, \dodoi{10.48550/arXiv.2412.01478}

\bibitem[{{Caleb} {et~al.}(2024){Caleb}, {Lenc}, {Kaplan}, {Murphy}, {Men},
  {Shannon}, {Ferrario}, {Rajwade}, {Clarke}, {Giacintucci}, {Hurley-Walker},
  {Hyman}, {Lower}, {McSweeney}, {Ravi}, {Barr}, {Buchner}, {Flynn}, {Hessels},
  {Kramer}, {Pritchard}, \& {Stappers}}]{2024NatAs...8.1159C}
{Caleb}, M., {Lenc}, E., {Kaplan}, D.~L., {et~al.} 2024, Nature Astronomy, 8,
  1159, \dodoi{10.1038/s41550-024-02277-w}

\bibitem[{{CASA Team} {et~al.}(2022){CASA Team}, {Bean}, {Bhatnagar}, {Castro},
  {Donovan Meyer}, {Emonts}, {Garcia}, {Garwood}, {Golap}, {Gonzalez Villalba},
  {Harris}, {Hayashi}, {Hoskins}, {Hsieh}, {Jagannathan}, {Kawasaki},
  {Keimpema}, {Kettenis}, {Lopez}, {Marvil}, {Masters}, {McNichols},
  {Mehringer}, {Miel}, {Moellenbrock}, {Montesino}, {Nakazato}, {Ott}, {Petry},
  {Pokorny}, {Raba}, {Rau}, {Schiebel}, {Schweighart}, {Sekhar}, {Shimada},
  {Small}, {Steeb}, {Sugimoto}, {Suoranta}, {Tsutsumi}, {van Bemmel},
  {Verkouter}, {Wells}, {Xiong}, {Szomoru}, {Griffith}, {Glendenning}, \&
  {Kern}}]{2022PASP..134k4501C}
{CASA Team}, {Bean}, B., {Bhatnagar}, S., {et~al.} 2022, \pasp, 134, 114501,
  \dodoi{10.1088/1538-3873/ac9642}

\bibitem[{{Chatterjee} {et~al.}(2017){Chatterjee}, {Law}, {Wharton},
  {Burke-Spolaor}, {Hessels}, {Bower}, {Cordes}, {Tendulkar}, {Bassa},
  {Demorest}, {Butler}, {Seymour}, {Scholz}, {Abruzzo}, {Bogdanov}, {Kaspi},
  {Keimpema}, {Lazio}, {Marcote}, {McLaughlin}, {Paragi}, {Ransom}, {Rupen},
  {Spitler}, \& {van Langevelde}}]{2017Natur.541...58C}
{Chatterjee}, S., {Law}, C.~J., {Wharton}, R.~S., {et~al.} 2017, \nat, 541, 58,
  \dodoi{10.1038/nature20797}

\bibitem[{{Chevalier}(1998)}]{1998ApJ...499..810C}
{Chevalier}, R.~A. 1998, \apj, 499, 810, \dodoi{10.1086/305676}

\bibitem[{{CHIME/FRB Collaboration} {et~al.}(2020){CHIME/FRB Collaboration},
  {Andersen}, {Bandura}, {Bhardwaj}, {Bij}, {Boyce}, {Boyle}, {Brar},
  {Cassanelli}, {Chawla}, {Chen}, {Cliche}, {Cook}, {Cubranic}, {Curtin},
  {Denman}, {Dobbs}, {Dong}, {Fandino}, {Fonseca}, {Gaensler}, {Giri}, {Good},
  {Halpern}, {Hill}, {Hinshaw}, {H{\"o}fer}, {Josephy}, {Kania}, {Kaspi},
  {Landecker}, {Leung}, {Li}, {Lin}, {Masui}, {McKinven}, {Mena-Parra},
  {Merryfield}, {Meyers}, {Michilli}, {Milutinovic}, {Mirhosseini},
  {M{\"u}nchmeyer}, {Naidu}, {Newburgh}, {Ng}, {Patel}, {Pen},
  {Pinsonneault-Marotte}, {Pleunis}, {Quine}, {Rafiei-Ravandi}, {Rahman},
  {Ransom}, {Renard}, {Sanghavi}, {Scholz}, {Shaw}, {Shin}, {Siegel}, {Singh},
  {Smegal}, {Smith}, {Stairs}, {Tan}, {Tendulkar}, {Tretyakov}, {Vanderlinde},
  {Wang}, {Wulf}, \& {Zwaniga}}]{CHIME-SGR}
{CHIME/FRB Collaboration}, {Andersen}, B.~C., {Bandura}, K.~M., {et~al.} 2020,
  \nat, 587, 54, \dodoi{10.1038/s41586-020-2863-y}

\bibitem[{{Condon} \& {Ransom}(2016)}]{essential2016}
{Condon}, J.~J., \& {Ransom}, S.~M. 2016, {Essential Radio Astronomy}
  (Princeton University Press)

\bibitem[{{Cordes} \& {Lazio}(2002)}]{2002_Cordes1}
{Cordes}, J.~M., \& {Lazio}, T.~J.~W. 2002, arXiv e-prints, astro.
\newblock \doarXiv{astro-ph/0207156}

\bibitem[{{Cordes} \& {Lazio}(2003)}]{2003_Cordes2}
---. 2003, arXiv e-prints, astro.
\newblock \doarXiv{astro-ph/0301598}

\bibitem[{{Curtin} {et~al.}(2023){Curtin}, {Tendulkar}, {Josephy}, {Chawla},
  {Andersen}, {Kaspi}, {Bhardwaj}, {Cassanelli}, {Cook}, {Dong}, {Fonseca},
  {Gaensler}, {Kaczmarek}, {Lanmnan}, {Leung}, {Pearlman}, {Petroff},
  {Pleunis}, {Rafiei-Ravandi}, {Ransom}, {Shin}, {Scholz}, {Smith}, \&
  {Stairs}}]{curtin23}
{Curtin}, A.~P., {Tendulkar}, S.~P., {Josephy}, A., {et~al.} 2023, \apj, 954,
  154, \dodoi{10.3847/1538-4357/ace52f}

\bibitem[{{Eftekhari} {et~al.}(2019){Eftekhari}, {Berger}, {Margalit},
  {Blanchard}, {Patton}, {Demorest}, {Williams}, {Chatterjee}, {Cordes},
  {Lunnan}, {Metzger}, \& {Nicholl}}]{2019ApJ...876L..10E}
{Eftekhari}, T., {Berger}, E., {Margalit}, B., {et~al.} 2019, \apjl, 876, L10,
  \dodoi{10.3847/2041-8213/ab18a5}

\bibitem[{{Feng} {et~al.}(2022){Feng}, {Li}, {Yang}, {Zhang}, {Zhu}, {Zhang},
  {Lu}, {Wang}, {Dai}, {Lynch}, {Yao}, {Jiang}, {Niu}, {Zhou}, {Xu}, {Miao},
  {Niu}, {Meng}, {Qian}, {Tsai}, {Wang}, {Xue}, {Yue}, {Yuan}, {Zhang}, \&
  {Zhang}}]{2022Sci...375.1266F}
{Feng}, Y., {Li}, D., {Yang}, Y.-P., {et~al.} 2022, Science, 375, 1266,
  \dodoi{10.1126/science.abl7759}

\bibitem[{{Gal-Yam}(2012)}]{Gal-Yam2012Sci}
{Gal-Yam}, A. 2012, Science, 337, 927, \dodoi{10.1126/science.1203601}

\bibitem[{{Hewitt} {et~al.}(2024){Hewitt}, {Huang}, {Hessels}, {Cognard},
  {Guillemot}, {Ould-Boukattine}, {Snelders}, \&
  {Kirsten}}]{2024ATel16597....1H}
{Hewitt}, D.~M., {Huang}, J., {Hessels}, J.~W.~T., {et~al.} 2024, The
  Astronomer's Telegram, 16597, 1

\bibitem[{{Heywood}(2020)}]{2020ascl.soft09003H}
{Heywood}, I. 2020, {oxkat: Semi-automated imaging of MeerKAT observations},
  Astrophysics Source Code Library, record ascl:2009.003

\bibitem[{{Hugo} {et~al.}(2022){Hugo}, {Perkins}, {Merry}, {Mauch}, \&
  {Smirnov}}]{2022ASPC..532..541H}
{Hugo}, B.~V., {Perkins}, S., {Merry}, B., {Mauch}, T., \& {Smirnov}, O.~M.
  2022, in Astronomical Society of the Pacific Conference Series, Vol. 532,
  Astronomical Data Analysis Software and Systems XXX, ed. J.~E. {Ruiz},
  F.~{Pierfedereci}, \& P.~{Teuben}, 541, \dodoi{10.48550/arXiv.2206.09179}

\bibitem[{{Ibik} {et~al.}(2024){Ibik}, {Drout}, {Gaensler}, {Scholz},
  {Sridhar}, {Margalit}, {Clarke}, {Law}, {Tendulkar}, {Michilli}, {Eftekhari},
  {Bhardwaj}, {Burke-Spolaor}, {Chatterjee}, {Cook}, {Hessels}, {Kirsten},
  {Joseph}, {Kaspi}, {Lazda}, {Masui}, {Nimmo}, {Pandhi}, {Pearlman},
  {Pleunis}, {Rafiei-Ravandi}, {Shin}, \& {Smith}}]{2024ApJ...976..199I}
{Ibik}, A.~L., {Drout}, M.~R., {Gaensler}, B.~M., {et~al.} 2024, \apj, 976,
  199, \dodoi{10.3847/1538-4357/ad808e}

\bibitem[{{Kashiyama} \& {Murase}(2017)}]{2017ApJ...839L...3K}
{Kashiyama}, K., \& {Murase}, K. 2017, \apjl, 839, L3,
  \dodoi{10.3847/2041-8213/aa68e1}

\bibitem[{{Kenyon} {et~al.}(2018){Kenyon}, {Smirnov}, {Grobler}, \&
  {Perkins}}]{2018MNRAS.478.2399K}
{Kenyon}, J.~S., {Smirnov}, O.~M., {Grobler}, T.~L., \& {Perkins}, S.~J. 2018,
  \mnras, 478, 2399, \dodoi{10.1093/mnras/sty1221}

\bibitem[{{Kumar} {et~al.}(2024{\natexlab{a}}){Kumar}, {Maan}, \&
  {Bhusare}}]{2024ATel16452....1K}
{Kumar}, A., {Maan}, Y., \& {Bhusare}, Y. 2024{\natexlab{a}}, The Astronomer's
  Telegram, 16452, 1

\bibitem[{{Kumar} {et~al.}(2024{\natexlab{b}}){Kumar}, {Maan}, \&
  {Bhusare}}]{2024arXiv240612804K}
---. 2024{\natexlab{b}}, arXiv e-prints, arXiv:2406.12804,
  \dodoi{10.48550/arXiv.2406.12804}

\bibitem[{{Li} {et~al.}(2021){Li}, {Lin}, {Xiong}, {Ge}, {Li}, {Li}, {Lu},
  {Zhang}, {Tuo}, {Nang}, {Zhang}, {Xiao}, {Chen}, {Song}, {Xu}, {Liu}, {Jia},
  {Cao}, {Qu}, {Zhang}, {Gu}, {Liao}, {Zhao}, {Tan}, {Nie}, {Zhao}, {Zheng},
  {Zheng}, {Luo}, {Cai}, {Li}, {Xue}, {Bu}, {Chang}, {Chen}, {Chen}, {Chen},
  {Chen}, {Chen}, {Cui}, {Cui}, {Deng}, {Dong}, {Du}, {Fu}, {Gao}, {Gao},
  {Gao}, {Gu}, {Guan}, {Guo}, {Han}, {Huang}, {Huo}, {Jiang}, {Jiang}, {Jin},
  {Jin}, {Kong}, {Li}, {Li}, {Li}, {Li}, {Li}, {Li}, {Li}, {Liang}, {Liu},
  {Liu}, {Liu}, {Liu}, {Liu}, {Lu}, {Lu}, {Luo}, {Ma}, {Meng}, {Ou}, {Sai},
  {Shang}, {Song}, {Sun}, {Tao}, {Wang}, {Wang}, {Wang}, {Wang}, {Wang}, {Wen},
  {Wu}, {Wu}, {Wu}, {Xiao}, {Xu}, {Yang}, {Yang}, {Yang}, {Yang}, {Yi}, {Yin},
  {You}, {Zhang}, {Zhang}, {Zhang}, {Zhang}, {Zhang}, {Zhang}, {Zhang},
  {Zhang}, {Zhang}, {Zhou}, {Zhou}, {Zhu}, {Zhu}, \& {Zhuang}}]{HXMT-SGR}
{Li}, C.~K., {Lin}, L., {Xiong}, S.~L., {et~al.} 2021, Nature Astronomy, 5,
  378, \dodoi{10.1038/s41550-021-01302-6}

\bibitem[{{Limaye} \& {Spitler}(2024)}]{2024ATel16620....1L}
{Limaye}, P., \& {Spitler}, L. 2024, The Astronomer's Telegram, 16620, 1

\bibitem[{{Lorimer} {et~al.}(2007){Lorimer}, {Bailes}, {McLaughlin},
  {Narkevic}, \& {Crawford}}]{2007Sci...318..777L}
{Lorimer}, D.~R., {Bailes}, M., {McLaughlin}, M.~A., {Narkevic}, D.~J., \&
  {Crawford}, F. 2007, Science, 318, 777, \dodoi{10.1126/science.1147532}

\bibitem[{{Marcote} {et~al.}(2017){Marcote}, {Paragi}, {Hessels}, {Keimpema},
  {van Langevelde}, {Huang}, {Bassa}, {Bogdanov}, {Bower}, {Burke-Spolaor},
  {Butler}, {Campbell}, {Chatterjee}, {Cordes}, {Demorest}, {Garrett}, {Ghosh},
  {Kaspi}, {Law}, {Lazio}, {McLaughlin}, {Ransom}, {Salter}, {Scholz},
  {Seymour}, {Siemion}, {Spitler}, {Tendulkar}, \&
  {Wharton}}]{2017ApJ...834L...8M}
{Marcote}, B., {Paragi}, Z., {Hessels}, J.~W.~T., {et~al.} 2017, \apjl, 834,
  L8, \dodoi{10.3847/2041-8213/834/2/L8}

\bibitem[{{Margalit} \& {Metzger}(2018)}]{2018ApJ...868L...4M}
{Margalit}, B., \& {Metzger}, B.~D. 2018, \apjl, 868, L4,
  \dodoi{10.3847/2041-8213/aaedad}

\bibitem[{{McMullin} {et~al.}(2007){McMullin}, {Waters}, {Schiebel}, {Young},
  \& {Golap}}]{2007ASPC_McMullin}
{McMullin}, J.~P., {Waters}, B., {Schiebel}, D., {Young}, W., \& {Golap}, K.
  2007, in Astronomical Society of the Pacific Conference Series, Vol. 376,
  Astronomical Data Analysis Software and Systems XVI, ed. R.~A. {Shaw},
  F.~{Hill}, \& D.~J. {Bell}, 127

\bibitem[{{Mereghetti} {et~al.}(2020){Mereghetti}, {Savchenko}, {Ferrigno},
  {G{\"o}tz}, {Rigoselli}, {Tiengo}, {Bazzano}, {Bozzo}, {Coleiro},
  {Courvoisier}, {Doyle}, {Goldwurm}, {Hanlon}, {Jourdain}, {von Kienlin},
  {Lutovinov}, {Martin-Carrillo}, {Molkov}, {Natalucci}, {Onori}, {Panessa},
  {Rodi}, {Rodriguez}, {S{\'a}nchez-Fern{\'a}ndez}, {Sunyaev}, \&
  {Ubertini}}]{Integral-SGR}
{Mereghetti}, S., {Savchenko}, V., {Ferrigno}, C., {et~al.} 2020, \apjl, 898,
  L29, \dodoi{10.3847/2041-8213/aba2cf}

\bibitem[{{Metzger} {et~al.}(2017){Metzger}, {Berger}, \&
  {Margalit}}]{2017ApJ...841...14M}
{Metzger}, B.~D., {Berger}, E., \& {Margalit}, B. 2017, \apj, 841, 14,
  \dodoi{10.3847/1538-4357/aa633d}

\bibitem[{{Michilli} {et~al.}(2018){Michilli}, {Seymour}, {Hessels}, {Spitler},
  {Gajjar}, {Archibald}, {Bower}, {Chatterjee}, {Cordes}, {Gourdji}, {Heald},
  {Kaspi}, {Law}, {Sobey}, {Adams}, {Bassa}, {Bogdanov}, {Brinkman},
  {Demorest}, {Fernand ez}, {Hellbourg}, {Lazio}, {Lynch}, {Maddox}, {Marcote},
  {McLaughlin}, {Paragi}, {Ransom}, {Scholz}, {Siemion}, {Tendulkar}, {van
  Rooy}, {Wharton}, \& {Whitlow}}]{2018Natur.553..182M}
{Michilli}, D., {Seymour}, A., {Hessels}, J.~W.~T., {et~al.} 2018, \nat, 553,
  182, \dodoi{10.1038/nature25149}

\bibitem[{{Murase} {et~al.}(2016){Murase}, {Kashiyama}, \&
  {M{\'e}sz{\'a}ros}}]{2016MNRAS.461.1498M}
{Murase}, K., {Kashiyama}, K., \& {M{\'e}sz{\'a}ros}, P. 2016, \mnras, 461,
  1498, \dodoi{10.1093/mnras/stw1328}

\bibitem[{{Narayan}(1992)}]{1992_Narayan}
{Narayan}, R. 1992, Philosophical Transactions of the Royal Society of London
  Series A, 341, 151, \dodoi{10.1098/rsta.1992.0090}

\bibitem[{{Ng} \& {Romani}(2004)}]{2004ApJ...601..479N}
{Ng}, C.~Y., \& {Romani}, R.~W. 2004, \apj, 601, 479, \dodoi{10.1086/380486}

\bibitem[{{Niu} {et~al.}(2022){Niu}, {Aggarwal}, {Li}, {Zhang}, {Chatterjee},
  {Tsai}, {Yu}, {Law}, {Burke-Spolaor}, {Cordes}, {Zhang}, {Ocker}, {Yao},
  {Wan}, {Feng}, {Niino}, {Bochenek}, {Cruces}, {Connor}, {Jiang}, {Dai},
  {Luo}, {Li}, {Miao}, {Niu}, {Anna-Thomas}, {Sydnor}, {Stern}, {Wang}, {Yuan},
  {Yue}, {Zhou}, {Yan}, {Zhu}, \& {Zhang}}]{2022Nature_Niu}
{Niu}, C.~H., {Aggarwal}, K., {Li}, D., {et~al.} 2022, \nat, 606, 873,
  \dodoi{10.1038/s41586-022-04755-5}

\bibitem[{{Offringa} {et~al.}(2014){Offringa}, {McKinley}, {Hurley-Walker},
  {Briggs}, {Wayth}, {Kaplan}, {Bell}, {Feng}, {Neben}, {Hughes}, {Rhee},
  {Murphy}, {Bhat}, {Bernardi}, {Bowman}, {Cappallo}, {Corey}, {Deshpande},
  {Emrich}, {Ewall-Wice}, {Gaensler}, {Goeke}, {Greenhill}, {Hazelton},
  {Hindson}, {Johnston-Hollitt}, {Jacobs}, {Kasper}, {Kratzenberg}, {Lenc},
  {Lonsdale}, {Lynch}, {McWhirter}, {Mitchell}, {Morales}, {Morgan},
  {Kudryavtseva}, {Oberoi}, {Ord}, {Pindor}, {Procopio}, {Prabu}, {Riding},
  {Roshi}, {Shankar}, {Srivani}, {Subrahmanyan}, {Tingay}, {Waterson},
  {Webster}, {Whitney}, {Williams}, \& {Williams}}]{2014MNRAS.444..606O}
{Offringa}, A.~R., {McKinley}, B., {Hurley-Walker}, N., {et~al.} 2014, \mnras,
  444, 606, \dodoi{10.1093/mnras/stu1368}

\bibitem[{{Ould-Boukattine} {et~al.}(2024){Ould-Boukattine}, {Hessels},
  {Kirsten}, {Hewitt}, {Snelders}, {Blaauw}, {Sluman}, {Mulder}, {Herrmann},
  {Gawronski}, {Puchalska}, \& {Gopinath}}]{2024ATel16432....1O}
{Ould-Boukattine}, O.~S., {Hessels}, J.~W.~T., {Kirsten}, F., {et~al.} 2024,
  The Astronomer's Telegram, 16432, 1

\bibitem[{{Pacholczyk}(1970)}]{1970ranp.book.....P}
{Pacholczyk}, A.~G. 1970, {Radio astrophysics. Nonthermal processes in galactic
  and extragalactic sources} (W.H.Freeman \& Co Ltd)

\bibitem[{{Panda} {et~al.}(2024){Panda}, {Roy}, {Bhattacharyya}, {Dudeja}, \&
  {Kudale}}]{2024arXiv240509749P}
{Panda}, U., {Roy}, J., {Bhattacharyya}, S., {Dudeja}, C., \& {Kudale}, S.
  2024, arXiv e-prints, arXiv:2405.09749, \dodoi{10.48550/arXiv.2405.09749}

\bibitem[{{Pearlman} {et~al.}(2024){Pearlman}, {Scholz}, {Bethapudi},
  {Hessels}, {Kaspi}, {Kirsten}, {Nimmo}, {Spitler}, {Fonseca}, {Meyers},
  {Stairs}, {Tan}, {Bhardwaj}, {Chatterjee}, {Cook}, {Curtin}, {Dong},
  {Eftekhari}, {Gaensler}, {G{\"u}ver}, {Kaczmarek}, {Leung}, {Masui},
  {Michilli}, {Prince}, {Sand}, {Shin}, {Smith}, \& {Tendulkar}}]{pearlman24}
{Pearlman}, A.~B., {Scholz}, P., {Bethapudi}, S., {et~al.} 2024, Nature
  Astronomy, \dodoi{10.1038/s41550-024-02386-6}

\bibitem[{{Pietka} {et~al.}(2015){Pietka}, {Fender}, \&
  {Keane}}]{2015MNRAS.446.3687P}
{Pietka}, M., {Fender}, R.~P., \& {Keane}, E.~F. 2015, \mnras, 446, 3687,
  \dodoi{10.1093/mnras/stu2335}

\bibitem[{{Porth} {et~al.}(2014){Porth}, {Komissarov}, \&
  {Keppens}}]{2014MNRAS.438..278P}
{Porth}, O., {Komissarov}, S.~S., \& {Keppens}, R. 2014, \mnras, 438, 278,
  \dodoi{10.1093/mnras/stt2176}

\bibitem[{{Resmi} {et~al.}(2021){Resmi}, {Vink}, \&
  {Ishwara-Chandra}}]{2021A&A...655A.102R}
{Resmi}, L., {Vink}, J., \& {Ishwara-Chandra}, C.~H. 2021, \aap, 655, A102,
  \dodoi{10.1051/0004-6361/202039771}

\bibitem[{Rigault(2018)}]{mickael_rigault_2018}
Rigault, M. 2018, ztfquery, a python tool to access ZTF data, doi,  Zenodo,
  \dodoi{10.5281/zenodo.1345222}

\bibitem[{{Sarbadhicary} {et~al.}(2021){Sarbadhicary}, {Tremou}, {Stewart},
  {Chomiuk}, {Peters}, {Hales}, {Strader}, {Momjian}, {Fender}, \&
  {Wilcots}}]{2021ApJ_Sarbadhicary}
{Sarbadhicary}, S.~K., {Tremou}, E., {Stewart}, A.~J., {et~al.} 2021, \apj,
  923, 31, \dodoi{10.3847/1538-4357/ac2239}

\bibitem[{{Shin} \& {CHIME/FRB Collaboration}(2024)}]{2024ATel16420....1S}
{Shin}, K., \& {CHIME/FRB Collaboration}. 2024, The Astronomer's Telegram,
  16420, 1

\bibitem[{{Snelders} {et~al.}(2024){Snelders}, {Bhandari}, {Kirsten},
  {Hessels}, {Marcote}, {Hewitt}, {Gawronski}, {Puchalska}, {Ould-Boukattine},
  {Gopinath}, {Nimmo}, {Karuppusamy}, {Herrmann}, {Yang}, {Blaauw},
  {Buttaccio}, {Maccaferri}, {Bach}, {Feiler}, {Bray}, {Williams}, {Wrigley},
  {Keimpema}, {Paragi}, {Burgay}, {Corongiu}, {Giroletti}, {Kramer}, {Pilia},
  {Spitler}, {Surcis}, {Trudu}, {Yuan}, {Wang}, \&
  {Bezrukovs}}]{2024ATel16542....1S}
{Snelders}, M.~P., {Bhandari}, S., {Kirsten}, F., {et~al.} 2024, The
  Astronomer's Telegram, 16542, 1

\bibitem[{{Soderberg} {et~al.}(2012){Soderberg}, {Margutti}, {Zauderer},
  {Krauss}, {Katz}, {Chomiuk}, {Dittmann}, {Nakar}, {Sakamoto}, {Kawai},
  {Hurley}, {Barthelmy}, {Toizumi}, {Morii}, {Chevalier}, {Gurwell},
  {Petitpas}, {Rupen}, {Alexander}, {Levesque}, {Fransson}, {Brunthaler},
  {Bietenholz}, {Chugai}, {Grindlay}, {Copete}, {Connaughton}, {Briggs},
  {Meegan}, {von Kienlin}, {Zhang}, {Rau}, {Golenetskii}, {Mazets}, \&
  {Cline}}]{2012ApJ...752...78S}
{Soderberg}, A.~M., {Margutti}, R., {Zauderer}, B.~A., {et~al.} 2012, \apj,
  752, 78, \dodoi{10.1088/0004-637X/752/2/78}

\bibitem[{{Sridhar} \& {Metzger}(2022)}]{2022ApJ_Sridhar}
{Sridhar}, N., \& {Metzger}, B.~D. 2022, \apj, 937, 5,
  \dodoi{10.3847/1538-4357/ac8a4a}

\bibitem[{{Thornton} {et~al.}(2013){Thornton}, {Stappers}, {Bailes},
  {Barsdell}, {Bates}, {Bhat}, {Burgay}, {Burke-Spolaor}, {Champion}, {Coster},
  {D'Amico}, {Jameson}, {Johnston}, {Keith}, {Kramer}, {Levin}, {Milia}, {Ng},
  {Possenti}, \& {van Straten}}]{2013Sci...341...53T}
{Thornton}, D., {Stappers}, B., {Bailes}, M., {et~al.} 2013, Science, 341, 53,
  \dodoi{10.1126/science.1236789}

\bibitem[{{Tian} {et~al.}(2024{\natexlab{a}}){Tian}, {Pastor-Marazuela},
  {Stappers}, {Rajwade}, {Caleb}, {Bezuidenhout}, {Barr}, \&
  {Kramer}}]{2024ATel16446....1T}
{Tian}, J., {Pastor-Marazuela}, I., {Stappers}, B., {et~al.}
  2024{\natexlab{a}}, The Astronomer's Telegram, 16446, 1

\bibitem[{{Tian} {et~al.}(2024{\natexlab{b}}){Tian}, {Rajwade},
  {Pastor-Marazuela}, {Stappers}, {Bezuidenhout}, {Caleb}, {Jankowski}, {Barr},
  \& {Kramer}}]{2024MNRAS.533.3174T}
{Tian}, J., {Rajwade}, K.~M., {Pastor-Marazuela}, I., {et~al.}
  2024{\natexlab{b}}, \mnras, 533, 3174, \dodoi{10.1093/mnras/stae2013}

\bibitem[{{Verrecchia} {et~al.}(2024){Verrecchia}, {Perri}, {Tavani}, {Pilia},
  {Casentini}, {Pelliciari}, {Pittori}, {Bernardi}, {Bianchi}, {Naldi},
  {Pupillo}, {Geminardi}, {Esposito}, {Limaye}, \&
  {Spitler}}]{2024ATel16645....1V}
{Verrecchia}, F., {Perri}, M., {Tavani}, M., {et~al.} 2024, The Astronomer's
  Telegram, 16645, 1

\bibitem[{{Walker}(1998)}]{1998MNRAS.294..307W}
{Walker}, M.~A. 1998, \mnras, 294, 307,
  \dodoi{10.1046/j.1365-8711.1998.01238.x}

\bibitem[{{Xing} \& {Yu}(2024{\natexlab{a}})}]{2024ATel16594....1X}
{Xing}, Y., \& {Yu}, W. 2024{\natexlab{a}}, The Astronomer's Telegram, 16594, 1

\bibitem[{{Xing} \& {Yu}(2024{\natexlab{b}})}]{2024ATel16630....1X}
---. 2024{\natexlab{b}}, The Astronomer's Telegram, 16630, 1

\bibitem[{{Xing} {et~al.}(2024){Xing}, {Yu}, {Yan}, {Zhang}, \&
  {Zhang}}]{2024arXiv241106996X}
{Xing}, Y., {Yu}, W., {Yan}, Z., {Zhang}, X., \& {Zhang}, B. 2024, arXiv
  e-prints, arXiv:2411.06996.
\newblock \doarXiv{2411.06996}

\bibitem[{{Yang} {et~al.}(2024){Yang}, {Feng}, {Tsai}, {Li}, {Shi}, {Wang},
  {Yang}, {Zhang}, {Niu}, {Yao}, {Cui}, {Su}, {Li}, {Zhang}, {Zhu}, \&
  {Cotton}}]{2024ApJ...976..165Y}
{Yang}, A.~Y., {Feng}, Y., {Tsai}, C.-W., {et~al.} 2024, \apj, 976, 165,
  \dodoi{10.3847/1538-4357/ad7d02}

\bibitem[{{Yang} {et~al.}(2020){Yang}, {Li}, \& {Zhang}}]{2020ApJ...895....7Y}
{Yang}, Y.-P., {Li}, Q.-C., \& {Zhang}, B. 2020, \apj, 895, 7,
  \dodoi{10.3847/1538-4357/ab88ab}

\bibitem[{{Yang} {et~al.}(2022){Yang}, {Lu}, {Feng}, {Zhang}, \&
  {Li}}]{2022ApJ...928L..16Y}
{Yang}, Y.-P., {Lu}, W., {Feng}, Y., {Zhang}, B., \& {Li}, D. 2022, \apjl, 928,
  L16, \dodoi{10.3847/2041-8213/ac5f46}

\bibitem[{{Yang} {et~al.}(2023){Yang}, {Xu}, \& {Zhang}}]{yang23}
{Yang}, Y.-P., {Xu}, S., \& {Zhang}, B. 2023, \mnras, 520, 2039,
  \dodoi{10.1093/mnras/stad168}

\bibitem[{{Yang} {et~al.}(2016){Yang}, {Zhang}, \& {Dai}}]{2016ApJ...819L..12Y}
{Yang}, Y.-P., {Zhang}, B., \& {Dai}, Z.-G. 2016, \apjl, 819, L12,
  \dodoi{10.3847/2041-8205/819/1/L12}

\bibitem[{{Zhang}(2023)}]{zhang23}
{Zhang}, B. 2023, Reviews of Modern Physics, 95, 035005,
  \dodoi{10.1103/RevModPhys.95.035005}

\bibitem[{{Zhang}(2024)}]{zhang24}
---. 2024, Annual Review of Nuclear and Particle Science, 74, 89,
  \dodoi{10.1146/annurev-nucl-102020-124444}

\bibitem[{{Zhang} {et~al.}(2024){Zhang}, {Wu}, {Cao}, {Zhu}, {Zhang}, {Niu},
  {Xie}, {Zhou}, {Wang}, {Zhu}, {Zhang}, {Wang}, {Niu}, {Di Li}, {Han}, {Lee},
  {Wang}, {Gao}, {Feng}, {Jiang}, {Jing}, {Li}, {Lu}, {Luo}, {Lyu}, {Wang},
  {Xu}, {Yang}, {Yu}, {Zhang}, \& {Project}}]{2024ATel16505....1Z}
{Zhang}, J., {Wu}, Q., {Cao}, S., {et~al.} 2024, The Astronomer's Telegram,
  16505, 1

\bibitem[{{Zhang} \& {Yu}(2024)}]{2024ATel16695....1Z}
{Zhang}, X., \& {Yu}, W. 2024, The Astronomer's Telegram, 16695, 1

\bibitem[{{Zhang} {et~al.}(2023){Zhang}, {Yu}, {Law}, {Li}, {Chatterjee},
  {Demorest}, {Yan}, {Niu}, {Aggarwal}, {Anna-Thomas}, {Burke-Spolaor},
  {Connor}, {Tsai}, {Zhu}, \& {Luo}}]{2023ApJ...959...89Z}
{Zhang}, X., {Yu}, W., {Law}, C., {et~al.} 2023, \apj, 959, 89,
  \dodoi{10.3847/1538-4357/ad0545}

\bibitem[{{Zhao} \& {Wang}(2021)}]{2021ApJ...923L..17Z}
{Zhao}, Z.~Y., \& {Wang}, F.~Y. 2021, \apjl, 923, L17,
  \dodoi{10.3847/2041-8213/ac3f2f}

\end{thebibliography}
\bibliographystyle{aasjournal}



\end{document}